\documentclass[apj,twocolappendix]{emulateapj-rtx4}
\usepackage{graphicx}

\newcommand{\myemail}{m.kuzuhara@nao.ac.jp}

\slugcomment{Submitted to ApJ on February 20th, 2013. Accepted July 8th, 2013.}

\shorttitle{Direct Imaging of GJ 504 b}

\shortauthors{Kuzuhara et al.}

\begin{document}

\title{Direct Imaging of a Cold Jovian Exoplanet in Orbit around the Sun-like Star GJ 504}

\author{M. Kuzuhara\altaffilmark{1, 2, 28}, M. Tamura\altaffilmark{2,3,13}, T. Kudo\altaffilmark{4}, M. Janson\altaffilmark{5}, R. Kandori\altaffilmark{2}, T. D. Brandt\altaffilmark{5}, C. Thalmann\altaffilmark{6}, D. Spiegel\altaffilmark{5,7},  B. Biller\altaffilmark{8}, J. Carson\altaffilmark{9}, Y. Hori\altaffilmark{2}, R. Suzuki\altaffilmark{2}, A. Burrows\altaffilmark{5}, T. Henning\altaffilmark{8}, E. L. Turner\altaffilmark{5,10}, M. W. McElwain\altaffilmark{11}, A. Moro-Mart{\'{\i}}n\altaffilmark{5,12}, T. Suenaga\altaffilmark{2, 3}, Y. H. Takahashi\altaffilmark{2, 13}, J. Kwon\altaffilmark{2,3}, P. Lucas\altaffilmark{14}, L. Abe\altaffilmark{15}, W. Brandner\altaffilmark{8}, S. Egner\altaffilmark{4}, M. Feldt\altaffilmark{8}, H. Fujiwara\altaffilmark{4}, M. Goto\altaffilmark{16}, C. A. Grady\altaffilmark{17, 11, 18}, O. Guyon\altaffilmark{4}, J. Hashimoto\altaffilmark{26}, Y. Hayano\altaffilmark{4}, M. Hayashi\altaffilmark{2}, S. S. Hayashi\altaffilmark{4}, K. W. Hodapp\altaffilmark{19}, M. Ishii\altaffilmark{4}, M. Iye\altaffilmark{2}, G. R. Knapp\altaffilmark{5}, T. Matsuo\altaffilmark{20}, S. Mayama\altaffilmark{21}, S. Miyama\altaffilmark{22}, J.-I. Morino\altaffilmark{2}, J. Nishikawa\altaffilmark{2,3},T. Nishimura\altaffilmark{4}, T. Kotani\altaffilmark{2}, N. Kusakabe\altaffilmark{2}, T. -S. Pyo\altaffilmark{4}, E. Serabyn\altaffilmark{23}, H. Suto\altaffilmark{2}, M. Takami\altaffilmark{24}, N. Takato\altaffilmark{4},H. Terada\altaffilmark{4}, D. Tomono\altaffilmark{4}, M. Watanabe\altaffilmark{25}, J. P. Wisniewski\altaffilmark{26},T. Yamada\altaffilmark{27}, H. Takami\altaffilmark{2}, T. Usuda\altaffilmark{2}}

\altaffiltext{1}{Department of Earth and Planetary Science, The University of Tokyo, 7-3-1 Hongo, Bunkyo-ku Tokyo, 113-0033, Japan}
\altaffiltext{2}{National Astronomical Observatory of Japan, 2-21-1 Osawa, Mitaka, Tokyo 181-8588, Japan; \myemail}
\altaffiltext{3}{Department of Astronomical Science, The Graduate University for Advanced Studies, 2-21-1 Osawa, Mitaka, Tokyo 181-8588, Japan}
\altaffiltext{4}{Subaru Telescope, National Astronomical Observatory of Japan, 650 North A`ohoku Place, Hilo, HI96720, USA}
\altaffiltext{5}{Department of Astrophysical Sciences, Princeton University, Peyton Hall, Ivy Lane, Princeton, NJ08544, USA}
\altaffiltext{6}{Astronomical Institute ``Anton Pannekoek'', University of Amsterdam, Postbus 94249, 1090 GE, Amsterdam, The Netherlands}
\altaffiltext{7}{Institute for Advanced Study, Princeton, NJ 08540, USA}
\altaffiltext{8}{Max Planck Institute for Astronomy, K\"onigstuhl 17, 69117 Heidelberg, Germany}
\altaffiltext{9}{Department of Physics and Astronomy, College of Charleston, 58 Coming St., Charleston, SC 29424, USA}
\altaffiltext{10}{Kavli Institute for Physics and Mathematics of the Universe, The University of Tokyo, 5-1-5, Kashiwanoha, Kashiwa, Chiba 277-8568, Japan}
\altaffiltext{11}{Exoplanets and Stellar Astrophysics Laboratory, Code 667, Goddard Space Flight Center, Greenbelt, MD 20771 USA}
\altaffiltext{12}{Department of Astrophysics, CAB-CSIC/INTA, 28850 Torrej\'on de Ardoz, Madrid, Spain}
\altaffiltext{13}{Department of Astronomy, The University of Tokyo, 7-3-1, Hongo, Bunkyo-ku, Tokyo 113-0033, Japan}
\altaffiltext{14}{Centre for Astrophysics, University of Hertfordshire, College Lane, Hatfield AL10 9AB, UK}
\altaffiltext{15}{Laboratoire Lagrange (UMR 7293), Universit\'e de Nice-Sophia Antipolis, CNRS, Observatoire de la C\^ote d'Azur, 28 avenue Valrose, 06108 Nice Cedex 2, France}
\altaffiltext{16}{Universit\"ats-Sternwarte M\"unchen, Ludwig-Maximilians-Universit\"at, Scheinerstr. 1, 81679 M\"unchen, Germany}
\altaffiltext{17}{Eureka Scientific, 2452 Delmer, Suite 100, Oakland CA 96002, USA}
\altaffiltext{18}{Goddard Center for Astrobiology}
\altaffiltext{19}{Institute for Astronomy, University of Hawaii, 640 N. A`ohoku Place, Hilo, HI 96720, USA}
\altaffiltext{20}{Department of Astronomy, Kyoto University, Kitashirakawa-Oiwake-cho, Sakyo-ku, Kyoto, Kyoto 606-8502, Japan}
\altaffiltext{21}{The Graduate University for Advanced Studies, Shonan International Village, Hayama, Miura, Kanagawa 240-0193, Japan}
\altaffiltext{22}{Hiroshima University, 1-3-2, Kagamiyama, Higashihiroshima, Hiroshima 739-8511, Japan}
\altaffiltext{23}{Jet Propulsion Laboratory, California Institute of Technology, Pasadena, CA 91109, USA}
\altaffiltext{24}{Institute of Astronomy and Astrophysics, Academia Sinica, P.O. Box 23-141, Taipei 10617, Taiwan}
\altaffiltext{25}{Department of Cosmosciences, Hokkaido University, Kita-ku, Sapporo, Hokkaido 060-0810, Japan}
\altaffiltext{26}{H.L. Dodge Department of Physics \& Astronomy, University of Oklahoma, 440 W Brooks St Norman, OK 73019, USA}
\altaffiltext{27}{Astronomical Institute, Tohoku University, Aoba-ku, Sendai, Miyagi 980-8578, Japan}
\altaffiltext{28}{Department of Earth and Planetary Sciences, Tokyo Institute of Technology, 2-12-1 Ookayama, Meguro-ku, Tokyo 152-8551, Japan}

\begin{abstract}

Several exoplanets have recently been imaged at wide separations of $>$10 AU from their parent stars. These span a limited range of ages ($<$50 Myr) and atmospheric properties, with temperatures of 800--1800 K and very red colors ($J - H > 0.5$ mag), implying thick cloud covers. Furthermore, substantial model uncertainties exist at these young ages due to the unknown initial conditions at formation, which can lead to an order of magnitude of uncertainty in the modeled planet mass. Here, we report the direct imaging discovery of a Jovian exoplanet around the Sun-like star GJ 504, detected as part of the SEEDS survey. The system is older than all other known directly-imaged planets; as a result, its estimated mass remains in the planetary regime independent of uncertainties related to choices of initial conditions in the exoplanet modeling. Using the most common exoplanet cooling model, and given the system age of 160$^{+350}_{-60}$ Myr, GJ 504 b has an estimated mass of 4$^{+4.5}_{-1.0}$ Jupiter masses, among the lowest of directly imaged planets. Its projected separation of 43.5 AU exceeds the typical outer boundary of $\sim$30 AU predicted for the core accretion mechanism. GJ 504 b is also significantly cooler (510$^{+30}_{-20}$ K) and has a bluer color ($J-H = -0.23$ mag) than previously imaged exoplanets, suggesting a largely cloud-free atmosphere accessible to spectroscopic characterization. Thus, it has the potential of providing novel insights into the origins of giant planets, as well as their atmospheric properties.

\end{abstract}


\section{Introduction}
More than 890 extrasolar planets (or exoplanets) have been discovered by now \citep{Schneider2011Catalog}, and NASA's {\it Kepler} mission alone has recently added more than 2,000 exoplanet candidates \citep{Batalha2012Kepler}. The vast majority of exoplanets have been discovered using indirect detection techniques that infer the presence of the exoplanet by monitoring the host star, such as planetary transits or radial-velocity variations. The techniques require observations over at least one orbital period, making them impractical for detecting long-period exoplanets at large separations. Meanwhile, in spite of the technical challenges, 8-m class telescopes equipped with adaptive optics systems and/or coronagraphs have recently reached the high-contrast performance necessary to image massive planets at large orbital separations \citep{Marois2008HR8799,Marois2010HR8799,Lagrange2010BetaPic}. Direct imaging surveys using such instruments provide measurements of an exoplanet's position and luminosity, which can be used to estimate its mass using the age of its host star and models of luminosity evolution for planetary-mass objects. \par

Because massive exoplanets cool and fade with time, direct-imaging searches have been most successful around young stars. Mass estimates for young planets depend strongly on the assumed system age, planetary atmosphere models, and initial thermodynamic state, leading to large uncertainties in the inferred mass. Planetary mass estimates conventionally assume a ``hot start'', in which the planet is initially in a high-temperature, high-entropy (and hence, luminous) state \citep{Baraffe2003COND, Burrows1997}. However, recent theoretical models suggest that giant planets produced according to standard formation theories could initially be much colder \citep{Marley2007hotstart, Fortney2008hotstart, Spiegel2012ColdStart}. The difference between the mass-luminosity relations of the ``hot-start'' and ``cold-start'' models decreases with exoplanet age; the models converge after $\sim$100 Myr for a 5 $M_{\rm{Jup}}$ planet, and $\sim$1 Gyr for a 10 $M_{\rm{Jup}}$ planet \citep{Marley2007hotstart, Fortney2008hotstart, Spiegel2012ColdStart}. \par

Previously imaged exoplanets are all younger than 50 Myr \citep[][]{Marois2008HR8799, Lagrange2010BetaPic, Carson2012}. The hot-start models imply planetary masses \citep{Marois2008HR8799,Marois2010HR8799,Lagrange2010BetaPic}, while cold-start models cannot reproduce the high observed luminosities \citep{Marley2007hotstart, Spiegel2012ColdStart}. Therefore, their planetary status would be negated if the cold-start models apply at such young ages. Nevertheless, independent constraints, such as dynamical stability analysis in the case of the HR 8799 system \citep[e.g.,][]{Fabrycky2010_HR8799,Marois2010HR8799,Sudol2012_HR8799}, in some cases can be used to exclude the coldest range of initial conditions, and confine their masses to the planetary-mass regime. In addition, we note that neither hot-start nor cold-start models have been calibrated against dynamical mass estimates. These arguments suggest that no fully established model for estimating the masses of directly imaged planets exists as of yet. However, direct imaging of an older system with a well-determined age could remove much of the model uncertainty related to the unknown initial conditions.  \par

A common feature of the previously imaged planets is large effective temperatures \citep[$>800$ K;][]{Marois2008HR8799}, and very red colors implying the presence of thick clouds in their atmospheres. Furthermore, they have generally been discovered as companions to massive ($>1.5$ $M_{\rm{\odot}}$) stars \citep[e.g.,][]{Baines2012HR8799Radius, Lagrange2010BetaPic, Carson2012}, hence they populate a limited range in atmospheric conditions and host star properties. Widening this range would greatly enhance our understanding of the population of exoplanets in wide orbits.  \par

Here, we report the detection of an exoplanet around the nearby Sun-like star GJ 504 using high-contrast near-infrared imaging on the Subaru Telescope. The projected separation of the planet (GJ 504 b) is measured to be 43.5 AU from the star, and the mass of planet (GJ 504 b) is estimated to be a few $M_{\rm{Jup}}$; because the system has an age of 160$^{+350}_{-60}$ Myr, its mass estimate is only weakly dependent on the uncertainty of initial conditions in the cooling models of giant planets. In Section 2, we describe the properties of GJ 504. In Section 3, the observations and data reductions are detailed. Section 4 presents a derivation of the mass of GJ 504 b and other interesting properties of this planet. Section 5 investigates the atmospheric properties and possible origin of GJ 504 b. Our results are summarized in Section 6.  

\section{GJ 504 Stellar Properties}

GJ 504 is an isolated G0-type main-sequence star. It has a mass of 1.2 $M_{\rm{\odot}}$ and an age of 160$^{+350}_{-60}$ Myr. In the following sections, we will describe the stellar properties in more detail, including its age estimate. Table \ref{table: system_property} summarizes the properties of GJ 504.

\begin{deluxetable*}{ccc}
\tabletypesize{\small}
\tablecaption{Properties of GJ 504 System}
\tablewidth{0pt}
\tablecolumns{4}
\tablehead{ \colhead{ Property} & \colhead{Primary} &\colhead{Planet}} 
\startdata
    Spectral type & G0V\tablenotemark{a} & late T--early Y \\
    Distance (pc) & \multicolumn{2}{c}{17.56 $\pm$ 0.08\tablenotemark{b}} \\
    Proper motion (mas yr$^{-1}$); $\rm{\mu_{\alpha}}$, $\rm{\mu_{\beta}}$ & \multicolumn{2}{c}{--338.83 $\pm$ 0.25, 190.24 $\pm$ 0.21\tablenotemark{b}} \\
    Effective temperature; $T_{\rm{eff}}$ (K)& 6234 $\pm$ 25\tablenotemark{c} & 510$^{+30}_{-20}$\tablenotemark{d}  \\
    Surface gravity; $\log{g}$ & 4.60 $\pm$ 0.02\tablenotemark{c} & 3.9$^{+0.4}_{-0.2}$\tablenotemark{d} \\
    Luminosity; $\log{(L/L_{\mathrm{\odot}})}$ & 0.332 $\pm$ 0.032\tablenotemark{c} &  --6.09$^{+0.06}_{-0.08}$\tablenotemark{d}\\
    Metallicity; [Fe/H] & 0.28 $\pm$ 0.03\tablenotemark{c} & $\cdots$ \\ 
    Adopted age (Myr) & \multicolumn{2}{c}{160$^{+350}_{-60}$} \\ 
    Age by gyrochronology (Myr) & \multicolumn{2}{c}{160$^{+70}_{-60}$} \\
    Age by chromospheric activity (Myr) & \multicolumn{2}{c}{330 $\pm$ 180} \\
    Mass & 1.22 $\pm$ 0.08 $M_{\rm{\odot}}$ &  4.0$^{+4.5}_{-1.0}$ $M_{\rm{Jup}}$\tablenotemark{d} \\
    $J$-band mass & $\cdots$ & 3.5$^{+3.5}_{-1.0}$ $M_{\rm{Jup}}$\\
    $H$-band mass & $\cdots$ & 3.0$^{+4.0}_{-0.5}$ $M_{\rm{Jup}}$\\
    $K_{\rm{s}}$-band mass  & $\cdots$ & 6.5$^{+5.0}_{-1.5}$ $M_{\rm{Jup}}$\\
    $L^{\rm{\prime}}$-band mass  & $\cdots$ & 3.5$^{+4.0}_{-1.5}$ $M_{\rm{Jup}}$\\
    NIR brightness\tablenotemark{e} (mag) \ ($J$; $\sim$1.2 $\rm{\mu}$m)  &  4.106 & 19.91 $\pm$ 0.15 \\
    \ \ \ \ \ \ \ \ \ \ \ \ \ \ \ \ \ \ \ \ \ \ \ \ \ \ \ \ \ \ \ ($H$; $\sim$1.6 $\rm{\mu}$m) & 3.860 & 20.14 $\pm$ 0.14\tablenotemark{f} \\
    \ \ \ \ \ \ \ \ \ \ \ \ \ \ \ \ \ \ \ \ \ \ \ \ \ \ \ \ \ \ \ ($K_{\rm{s}}$; $\sim$2.2 $\rm{\mu}$m) & 3.808 & 19.19 $\pm$ 0.28 \\   
    \ \ \ \ \ \ \ \ \ \ \ \ \ \ \ \ \ \ \ \ \ \ \ \ \ \ \ \ \ \ \ ($L^{\prime}$; $\sim$3.8 $\rm{\mu}$m) & 3.94 & 16.84 $\pm$  0.19
    \enddata
    \tablenotetext{a}{\citet{ValdesNSParameter} and \citet{Takeda2004NSParameter}.}
    \tablenotetext{b}{\citet{vanLeeuwen2007Hip}.}
    \tablenotetext{c}{\citet{Valenti2005_StellarParam}.}
    \tablenotetext{d}{GJ 504 b's effective temperature, surface gravity, bolometric luminosity, and mass are estimated by comparing its age (160$^{+350}_{-60}$ Myr) and $JHK_{\rm{s}}L^{\prime}$-band magnitude with the most commonly used cooling model \citep{Baraffe2003COND}. For quantifying best estimates and lower/upper limits for all of these quantities, we apply the same calculation as described in Section \ref{section: mass estimate}.}
    \tablenotetext{e}{The provided errors of the stellar $JHK$ photometry are less than 0.01 mag \citep{Kidger2003}, while $L^{\prime}$ photometry was obtained in this work with an error of 0.09 mag (see Section \ref{sec: photometry}).}
    \tablenotetext{f}{The $H$-band photometric magnitude is based on the weighted mean of the March and May values.}
 \label{table: system_property}
\end{deluxetable*}

\subsection{Age of GJ 504}
\begin{figure}
\epsscale{1.1}
\plotone{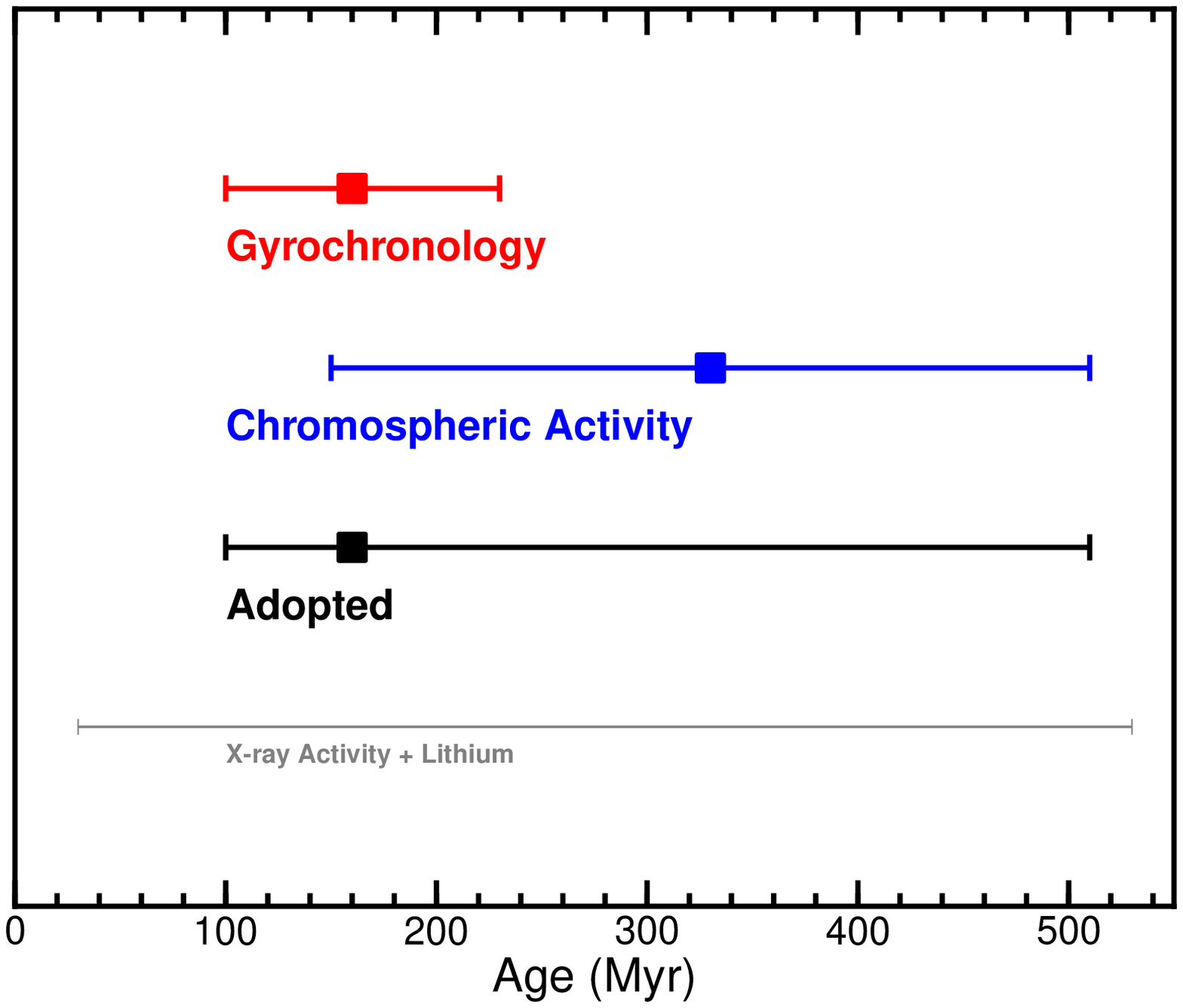}
\caption{GJ 504's ages independently inferred from gyrochronology, chromospheric and X-ray coronal activity, or lithium technique are summarized together with our adopted age for the system. The squares indicate the optimum estimates based on each technique, and the horizontal bars correspond to their error ranges. Because the gyrochronology and chromospheric activity could be the most reliable age indicators for isolated G-type stars with evolution stages near ZAMS, we use those as the fiducial age estimators. The best-estimate in our adopted age for GJ 504 is the gyrochronology technique, which directly traces the stellar rotation than the activity. The estimates from lithium or X-ray activity are not adopted but consistent to those from others.}
\label{fig: f0_age.eps}
\end{figure}

For an isolated main-sequence star like GJ 504, several methods of age estimation can be imagined, including gyrochronology, chromospheric or coronal activity, lithium abundance, kinematics, or isochrones. Among them, two particularly relevant age indicators for GJ 504 are gyrochronology based on stellar rotation, and chromospheric activity as traced by \ion{Ca}{2} H and K emission, and this study thus use these indicators as the fiducial age estimators. Both stellar rotation and activity decline as a Sun-like star ages and sheds angular momentum in its stellar wind, and both age indicators have been accurately calibrated using stars in open clusters \citep[][hereafter MH08]{MH08Age}. Chromospheric activity is powered by the stellar magnetic dynamo, which is closely related to the stellar rotation rate. Gyrochronology, which directly measures this rate, is therefore the most direct, and may be the most reliable, of our age estimators (MH08), and allows us to obtain a best estimate of 160$_{-60}^{+70}$ Myr for the age of GJ 504. However, adopting the age as given by the star's chromospheric activity in addition to that from gyrochronology, we conservatively adopt 160$^{+350}_{-60}$ Myr as the age estimate of GJ 504 system (see Figure \ref{fig: f0_age.eps} for summary). We describe these age estimations in detail below, along with several other age indicators that are less well suited for the purpose. 

\subsubsection{Gyrochronological Age} \label{sec: Gyro_age}

The rotation of a star slows as the star ages and the stellar wind carries away angular momentum. Gyrochronology, which estimates a star's age from its rotation rate and $B-V$ color (i.e., mass), is generally considered to be the most direct, and possibly the most accurate, method for inferring stellar age (MH08). Chromospheric and X-ray activity are powered by the stellar dynamo, which is closely related to the star's rotation rate. Indeed, chromospheric age estimators often use stellar activity to infer the rotation period, which is then used to infer an age (MH08). The rotation period of GJ 504 has been directly measured to be 3.328 days \citep{Messina2003ProtXray} and 3.33 days \citep{Donahue1996Prot}. We adopt the average of these measurements, 3.329 days, as the star's rotation period. The possible error range of $<$ 0.01 days is small enough not to affect the uncertainty of the age estimate. \par

The gyrochronology can be applied to I-sequence stars whose dynamo mechanism is presumed to originate at the interface between the convective and radiative zones in the stellar interior \citep{Barnes2007Gyro}. In contrast, this method cannot be applied to C-sequence stars that are fully convective and may thus have a different dynamo mechanism that prevents efficient spin-down. We can judge that GJ 504 is an I-sequence star using two different sets of criteria specified in the literature \citep{Barnes2007Gyro, Dupuy2009HD130948}: one in which we relate the age-normalized rotation of GJ 504 to other stars of similar $B-V$ colors, and the other in which the X-ray luminosity and Rossby number are compared to other chromospherically active stars. Both of these methods confirm that GJ 504 is consistent with an I-sequence star. This conclusion agrees with the previous classification \citep{Barnes2007Gyro} of GJ 504 as an I-sequence star.

To rigorously establish the age of GJ 504, we use three gyrochronology relations independently proposed \citep[MH08;][]{Barnes2007Gyro, Meibom2009Gyro}. All the relations determine a stellar age as a function of rotation period and $B-V$ color. Using the rotation period (3.329 days) and $B-V$ color (0.585 $\pm$ 0.007) of GJ 504, we derive ages of 140--180 (MH08), 100--140 \citep{Barnes2007Gyro}, and 110--230 Myr \citep{Meibom2009Gyro}, respectively. The error ranges for the age estimates were derived based on all the possible sources of error: the observed scatter in the period--age relation (0.05 dex; MH08) or errors of coefficients in the period--age function \citep{Barnes2007Gyro, Meibom2009Gyro}, and the error of the $B-V$ color. Conservatively adopting all these ranges,\footnote[1]{For the optimal value of the gyrochronological age estimation, we adopt 160 Myr which was calculated by MH08, because MH08 have updated the gyrochronological relation of \citet{Barnes2007Gyro}. Both relations of MH08 and \citet{Barnes2007Gyro} have been derived by utilizing several stellar clusters, while \citet{Meibom2009Gyro} used only a single cluster.} we estimate the age of GJ 504 to be 160$^{+70}_{-60}$ Myr. \par

\begin{figure}
\epsscale{1.18}
\plotone{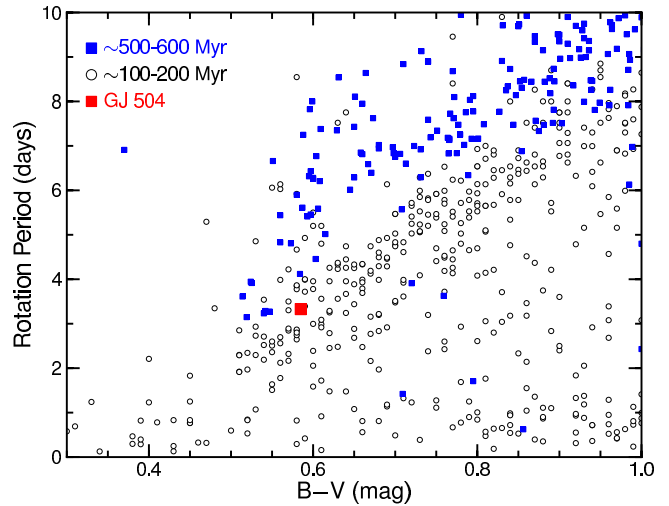}
\caption{Stellar rotation of GJ 504, and of stars with $B-V < 1.0$ in various clusters. The horizontal axis shows the $B-V$ colors of the stellar samples, while the vertical axis denotes their rotations. The data of stars in $\sim$100--200 Myr old clusters (Pleiades, M35, M34, and M11) are plotted as open circles, and those of $\sim$500--600 Myr old clusters (M37, Hyades, and Praesepe) are plotted as blue squares. See the main text for the references of these data. The uncertainties in stellar rotation periods should be smaller than $\sim$10\% in general \citep[cf.][]{Donahue1996Prot, Barnes2007Gyro, Hartman2009M37Rotation}. The squares with $B-V$ colors bluer than 0.55 mag are M37 stars, with uncertainties of $\sim$0.01 mag. Thus, GJ 504 does not overlap with stars in 500--600 Myr clusters.}
\label{fig: f1_rotations.pdf}
\end{figure}

In order to verify the adequacy of GK 504's age assessed from the empirical relations of gyrochronology (see above), we directly compare the rotation period and $B-V$ color of GJ 504 to those of stars in young clusters (see Figure \ref{fig: f1_rotations.pdf}). The rotation data for stellar members in Pleiades \citep[130 Myr;][]{Hartman2010PleiadesRotation}, M35 \citep[150 Myr;][]{Meibom2009Gyro}, M34 \citep[220 Myr;][]{Meibom2011Gyro}, M11 \citep[230 Myr;][]{Messina2010M11Rotation}, M37 \citep[550 Myr;][]{Hartman2009M37Rotation}, Hyades \citep[630 Myr;][]{Radick1987_Hyades_rotation,Radick1995_Hyades_rotation,Delorme2011_stellar_rot_Hyades_Praesepe}, and Praesepe \citep[600 Myr;][]{Delorme2011_stellar_rot_Hyades_Praesepe} can be used for this purpose. As shown in Figure \ref{fig: f1_rotations.pdf}, the stars with similar $B-V$ colors to GJ 504 in 130--230 Myr old clusters have periods that are the most consistent with that of GJ 504, and the rotation periods in the older 500--600 Myr cluster are significantly longer,\footnote[2]{In addition to I-sequence stars, Figure \ref{fig: f1_rotations.pdf} includes rotation data of C-sequence stars, which are widely scattered in the plot. However, the scatter is much smaller for the older clusters. GJ 504 is best represented by I-sequence stars, and young I-sequence stars such as provided in M35 or M34 \citep{Meibom2009Gyro,Meibom2011Gyro} are indeed in good agreement with GJ 504.} thus verifying the age range of GJ 504 derived above. Additionally, combining the rotation data of stars in M11 with those in M34, \citet{Messina2010M11Rotation} derive a median rotation period of G-type stars in these clusters of 4.8 days, again supporting the age estimate of 160$^{+70}_{-60}$ Myr for the G-type star GJ 504 (3.329 day period). \par

Since gyrochronology is a largely empirical method, there could in principle be uncertainties involved in the methodology that are not accounted for in the calibrations. For instance, if we hypothesize that the calibration could be metallicity-dependent, this would lead to systematic errors for targets with significantly different metallicity from the calibration clusters. For the case of GJ 504, it has a non-extreme value in this regard, hence it is unlikely to be subject to such uncertainties. Another possible systematic uncertainty in gyrochronology involves the fact that rotation is easier to detect in rapidly rotating systems, which could result in a bias in the empirical gyrochronology relation. If this were a significant effect, it would cause the apparent mean rotation in a cluster of a given age to appear more rapid than it is in reality. Hence, age estimates for individual stars based on a relation calibrated against such clusters would result in an overestimation (i.e., an individual star would appear older than it really is). Given that no other age indicators imply a significantly younger age for GJ 504 this is also unlikely to be a real issue, but nonetheless, such uncertainties are important to keep in mind in the age analysis.

\subsubsection{Chromospheric Activity Age} \label{sec: CA_age}

Stellar chromospheric activity, as traced by \ion{Ca}{2} H and K emission, is a good indicator of stellar age. The strength of \ion{Ca}{2} H and K emission is parameterized by $R'_{\rm{HK}}$, roughly the power in the H and K lines normalized to that in the underlying photospheric continuum. A relation between $R'_{\rm{HK}}$, $B-V$ color, and age has recently been derived for solar-type stars (MH08). For GJ 504, we adopt the value of log $R'_{\rm{HK}} = -4.45$ measured at Mt. Wilson Observatory with a temporal baseline of 30 yr \citep{Radick1998Prot}, the longest such baseline in the literature. In order to minimize the uncertainty in $R'_{\rm{HK}}$, the value for which the star has been monitored for the longest possible time should be adopted (see MH08), which also protects against accidentally adopting an outlier in the activity variation.\par

This activity level may be used to infer a Rossby number, the ratio of the stellar rotation period to the timescale of convective overturn, of 0.60 $\pm$ 0.10 following MH08.\footnote[3]{MH08 have constructed an empirical function to convert a measurement of $R'_{\rm{HK}}$ to the Rossby number. The error in the inferred Rossby number was measured to be 0.10 for the multi-decade Mount Wilson data (see Section 4.1.2 and summary of MH08).} We use this result,\footnote[4]{The Rossby number is converted into a rotation period, which is used in the gyrochronology relation (i.e., activity--rotation--age relation) to estimate the stellar age. This scheme improves the accuracy of age estimation from chromospheric activity (see MH08). } together with a $B-V$ color of 0.585 $\pm$ 0.007 \citep{vanLeeuwen2007Hip}, to estimate an age of 330 $\pm$ 100 Myr. We further add a 0.2 dex scatter in this relation, estimated from clusters\footnote[5]{MH08 infer the uncertainty of their activity-rotation-age relation by calculating the rms of residuals between the ages of clusters and those of their stellar members, which were estimated by applying the activity-rotation-age relation to the members.} (MH08). We assume this scatter to be uncorrelated with other sources of error. As a result, we obtain a chromospheric age estimate of 330 $\pm$ 180 Myr. \par

Other studies using data from Mt. Wilson Observatory have measured $R'_{\rm{HK}}$ values of --4.443 \citep{Baliunas1996Chromo} and --4.486 \citep{Lockwood2007ChHK}, which differ slightly from our adopted value of --4.45. The chromospheric activity level was also measured to be $R'_{\rm{HK}}=-4.44$ using 14 yr of observations at Fairborn and Lowell Observatories \citep{Hall2009CaHK}. Many of these studies' observational baselines overlap with the source of our adopted $R'_{\rm{HK}}$ value \citep{Radick1998Prot}. Adopting the mean of these measurements would have little effect on our chromospheric age of 330 $\pm$ 180 Myr. \par

A primary uncertainty in age determination from chromospheric activity is long-term activity cycles, which may easily exceed the timescales over which the activity is measured, and thus may potentially provide a systematic error if the activity is measured at an extreme but is taken to represent the mean activity for the star. In addition, as in the case of gyrochronology, the stellar activity also depends on the metallicity, possibly contributing to the uncertainty. Since activity as an age estimator is thought to be closely related to gyrochronology through the stellar dynamo, the possibility of measuring both rotation (which is robust against activity cycles) and activity in the same star as for the case of GJ 504 can alleviate this uncertainty.

\subsubsection{X-Ray Activity} \label{sec: Xray_age}

Young solar-type stars are active X-ray emitters. As for \ion{Ca}{2} H and K, a relation has been derived (MH08) between stellar age, color, and X-ray luminosity. The best estimate of the X-ray luminosity of GJ 504 is from the {\it ROSAT} all-sky survey of nearby stars \citep{Hunsch1999NSXray}, which has provided a ratio of X-ray to bolometric power of $\log{L_{\rm{X}}/L_{\rm{bol}}} = -4.42$ \citep{Messina2003ProtXray}. This X-ray activity may be used to infer a Rossby number of 0.54 $\pm$ 0.25 (MH08) and lead to the age estimate of 90--530 Myr for GJ 504. Because the temporal baseline for the X-ray measurement is much shorter than for the multi-decade measurements of $R'_{\rm{HK}}$, the inferred Rossby number is much more uncertain. Therefore, we do not adopt this estimate based on the X-ray emission, although it is consistent with the estimate from other methods.\par

\subsubsection{Lithium Absorption} \label{sec: Li_age}

The strength of lithium absorption lines declines as a star burns its primordial supply of lithium, providing another indicator of stellar youth  \citep{Soderbloom2010AgeReview}. However, many effects, such as variations in the primordial supply of lithium, the extreme sensitivity of early lithium burning to the stellar accretion history \citep{Baraffe2010EpisodicLi}, and the importance of non-local thermodynamic equilibrium abundance corrections \citep{Takeda2005Li}, make it difficult to obtain a reliable age estimate from lithium abundances. Therefore, we do not employ lithium as an indicator to date GJ 504, but it can provide a clue to the age of GJ 504 as discussed below. The equivalent width of Li in GJ 504 has been measured to be 0.08 \AA, implying a lithium abundance of $\log{n}$(Li) = $\log{(N(\rm{Li})/{\it N}(\rm{H}))}$ + 12 = 2.9 \citep{Takeda2005Li}. Extensive observations of well-dated open clusters \citep{Sestito2005Li} enable a lithium age estimate for GJ 504. At the effective temperature of a G0 star ($\sim$6000 K), these cluster estimates can be used to infer an age range of approximately 30--500 Myr, fully consistent with our gyrochronological and chromospheric age estimates. \par

\subsubsection{Kinematics} \label{sec: Kin_age}

GJ 504 is not currently identified as a member of any clusters or moving groups (MGs), making it impossible to use population-based age estimators. The kinematic velocity of GJ 504 \citep[$U$=--38.6, $V$=1.5, $W$=--18.0;][]{Mishenina2004Kinematics} is consistent with thin disk stars, thus providing no tight constraint on the age, but remaining consistent with the young age provided by gyrochronology and chromospheric activity. \par

\subsubsection{Isochrones} \label{sec: Iso_age}

Since stars evolve along mass tracks in an HR-diagram with age, a common age determination method is based on isochronal analysis. This method can provide reasonable estimates for old stars that have evolved significantly off the main sequence and very young stars that are still undergoing rapid contraction. However, it is generally not useful for stars near the zero-age main sequence (ZAMS), around which stellar evolution proceeds slowly \citep[e.g.,][]{Soderbloom2010AgeReview}. Indeed, generic isochronal estimates in the literature are known to give erroneous ages for young stars by up to two orders of magnitudes. \par

As illustrative examples, we consider the cases of HD~152555, HIP~10679, and HD~206860. These are all early G-type (G0--G2) stars which are members of young MGs, hence they are similar types of stars to GJ~504 with well-determined young ages, and thus can be used for comparison to isochronal ages of e.g., \citet{Valenti2005_StellarParam}. HD~152555 is a member of the AB~Dor MG \citep[e.g.,][]{Torres2008MGsummary} with an age of 50--100~Myr, whereas \citet{Valenti2005_StellarParam} give an age of 4.0~Gyr. HIP~10679 is a member of the $\beta$~Pic MG \citep[e.g.,][]{Torres2008MGsummary} with an age of $\sim$10~Myr, whereas \citet{Valenti2005_StellarParam} give an age of 3.0~Gyr. Finally, HD~206860 is a member of the Her-Lya MG \citep[e.g.,][]{Lopez-Santiago2006HerLya} with an age of $\sim$200~Myr, whereas \citet{Valenti2005_StellarParam} give an age of 3.1~Gyr. \par

\begin{figure*}
\plottwo{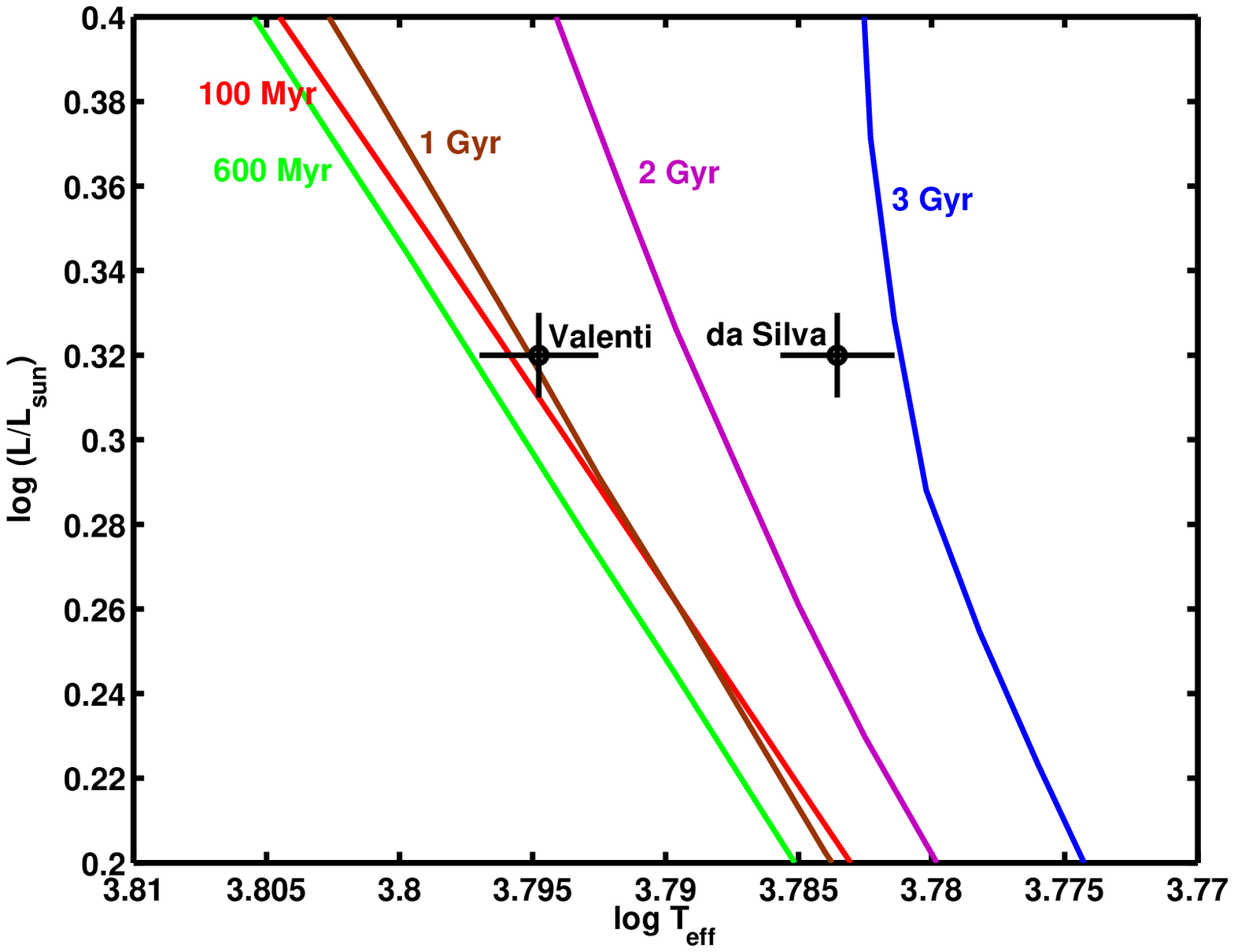}{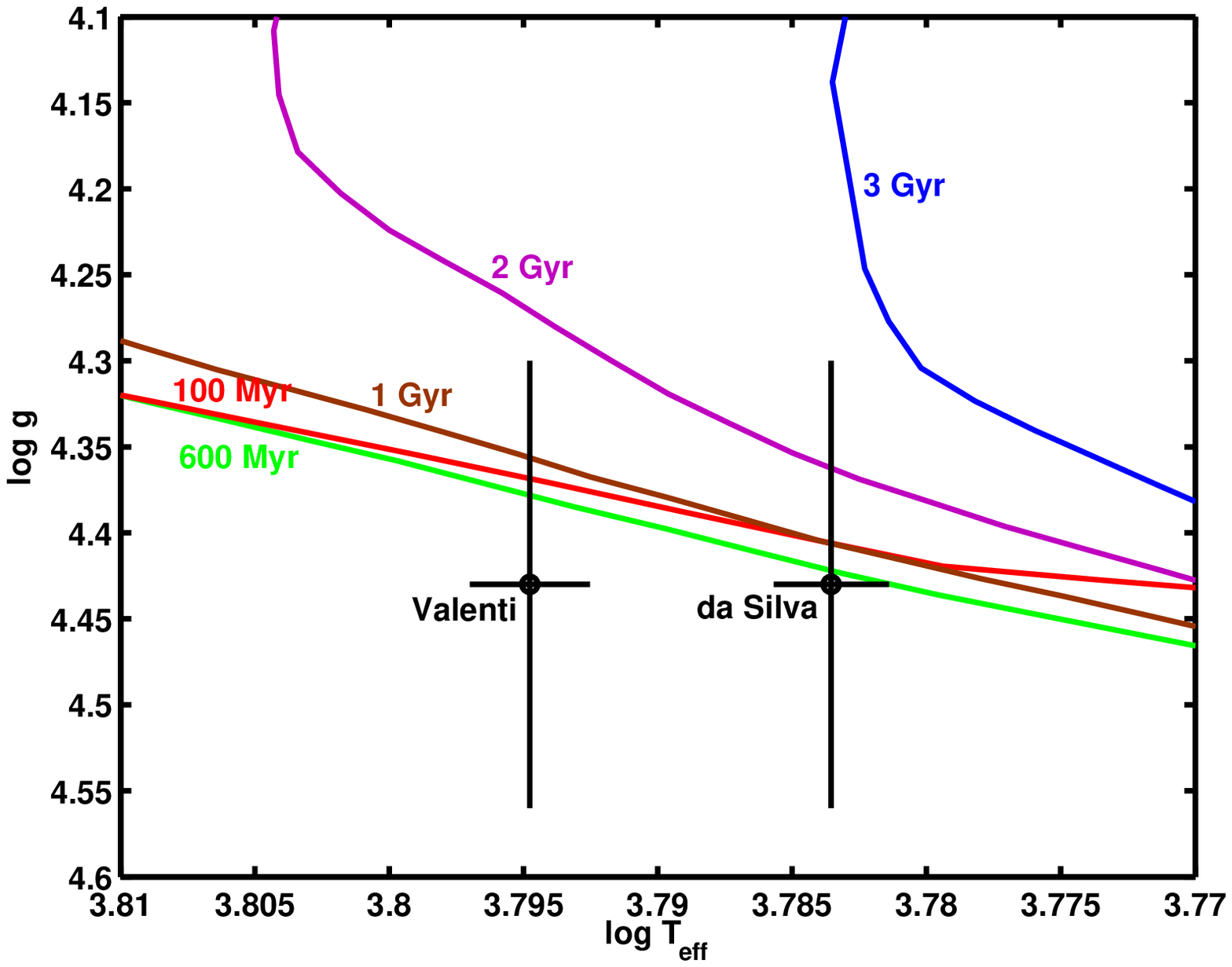}
\epsscale{1.3}
\caption{Isochronal analysis of GJ~504. Left: reproduction of the \citet{daSilva2012Age} result. Also plotted is another data point where all values are kept equal, except for the temperature which has been replaced with the one in \citet{Valenti2005_StellarParam}, illustrating the real uncertainty in temperature. With the \citet{Valenti2005_StellarParam} temperature, the data are fully consistent with the ages derived from gyrochronology and activity. Right: the same analysis, again with all quantities except for temperature fixed to the \citet{daSilva2012Age} values, but replacing luminosity with surface gravity. In this framework, $>$3 Gyr ages are disfavored regardless of which temperature is chosen.}
\label{f:isochrones}
\end{figure*}

In light of these large discrepancies, it is unsurprising that literature estimates for the isochronal age of GJ~504 vary strongly from hundreds of Myrs to several Gyrs \citep[e.g.,][]{Valenti2005_StellarParam,Takeda2007NSStructure,Holmberg2009Geneva,daSilva2012Age}. This implies that isochronal analysis should simply be disregarded for a star such as GJ~504, compared to the more reliable age estimators listed above. Nonetheless, since some of the isochronal literature estimates provide Gyr-range ages that are seemingly inconsistent with the younger age range we have adopted for GJ~504, we discuss their application to this star in more detail below. \par

An important limiting factor in isochrone analysis for age determination of an isolated main-sequence star such as GJ~504 is the sensitivity to uncertainties in input stellar parameters. This is in particular true for properties such as effective temperature and surface gravity, where the uncertainties are model-dependent and often tend to be under-estimated in the literature. As an illustration, we consider the isochronal analysis in \citet{daSilva2012Age} based on the Yonsei-Yale isochrones \citep{Yi2001YaleYonsei}, which gives an age of $\sim$3~Gyr, and appears to be inconsistent with ages of $\leq$500~Myr within their quoted error bars. We have reproduced the analysis of \citet{daSilva2012Age} and confirmed these results with the same methodology and input observables. This is plotted in Figure \ref{f:isochrones}. Also plotted in the same figure is the exact same analysis, but using the temperature estimates of GJ~504 from \citet{Valenti2005_StellarParam} instead; note that all other values are kept as in \citet{daSilva2012Age}. This comparison immediately illuminates one of the considerable uncertainties related to isochronal analysis, which is uncertainties in the temperature scaling. The two temperature estimates in \citet{Valenti2005_StellarParam} and \citet{daSilva2012Age} are inconsistent within their quoted error bars, implying that neither error estimate should be trusted when translating this into an error on the age from the isochronal comparison. Moreover, we find that while the temperature estimate of \citet{daSilva2012Age} gives the aforementioned age of $\sim$3~Gyr and is inconsistent with $\leq$500~Myr ages, the temperature estimate of \citet{Valenti2005_StellarParam} is fully consistent with these younger ages, and inconsistent with the $\sim$3~Gyr age. \par

It is also important to note that due to pre-main sequence evolution, there is some overlap between the isochrones of stars arriving at the main sequence, and those leaving it. For instance, the 100~Myr isochrone overlaps with the 1~Gyr isochrone in the relevant temperature range. Hence, while the age is quoted as $\sim$1.4~Gyr in \citet{Valenti2005_StellarParam} which is reproduced in Figure \ref{f:isochrones}, it can be equally well interpreted as being consistent with our adopted age range, and in fact preferentially supporting the youngest ages. \par

Furthermore, while the temperature and luminosity in \citet{daSilva2012Age} gives a certain age from comparison with the Yonsei-Yale isochrones, this does not account for the relatively high surface gravity they quote. We can see this in the second panel of Figure \ref{f:isochrones}, where again we perform an isochronal analysis, but in surface gravity versus temperature as opposed to luminosity versus temperature. Here, regardless of which temperature value used, the data are consistent with a $\leq$500~Myr age, but inconsistent with a $\sim$3~Gyr age. \par

It is instructive to think of these discrepancies in terms of the radius of the star. If we adopt the \citet{daSilva2012Age} temperature, then the luminosity is high for its mass, implying a large radius. On the other hand, the surface gravity is also high for its mass, implying a small radius. Hence, there is a tension in the radius between the luminosity and surface gravity if the \citet{daSilva2012Age} temperature is adopted, such that no self-consistent picture can be reached by either increasing or decreasing the stellar radius. By contrast, if we adopt the \citet{Valenti2005_StellarParam} temperature, then this tension disappears, and a self-consistent picture can be reached between the temperature, luminosity, and surface gravity---and simultaneously, we get a predicted age which is consistent with the estimates given by the age estimates in the previous sections. This could therefore be interpreted as supporting the higher temperature of \citet{Valenti2005_StellarParam} and thus a younger age. \par

Hence, we note that the isochronal age is readily consistent with gyrochronology and activity, using values for temperature, luminosity, and surface gravity published in the literature. This is also consistent with the arguably most detailed Bayesian isochronal analysis of \citet{Takeda2007NSStructure}, which was a study meant to improve on the analysis of \citet{Valenti2005_StellarParam} using the same input data, and which gives an age limit of $<$760~Myr. Nonetheless, the above discussions imply that it is not feasible to provide age constraints from isochronal dating that are as meaningful nor as stringent as those of gyrochronology and chromospheric activity. Thus, we do not include it in our adopted age range, but merely conclude that it is broadly consistent with such an age. \par

Finally, it can be noted that the existence of a star with such a short rotation period as GJ 504 ($\sim$3 days) would be unprecedented at the Gyr-scale ages predicted by some isochrone analyses, such as \citet{daSilva2012Age}. Indeed, although \citet{Reiners2009M67rotation} found an unusually rapid rotator in the $\sim$4 Gyr old M67 cluster, it has a rotation of approximately half the solar rotation rate ($P_{\rm{\odot}} \sim 30$ days), which is significantly slower than the rotation of GJ 504. Aside from known close spectroscopic binaries, it is the largest known outlier at such an age. No indications of chromospheric activity corresponding to such rapid rotation have been found among 60 stars in the M67 cluster \citep{Giampapa2006}, which supports the notion that rapid rotators like GJ 504 must be exceedingly rare at ages of a few Gyr. 

\subsubsection{Summary of Age Estimate for GJ 504}

As discussed above, we adopt the age estimated with gyrochronology as the most likely age (160$^{+70}_{-60}$ Myr) for GJ 504. However, we conservatively adopt an age of 160$^{+350}_{-60}$ Myr, with chromospheric activity providing the upper bound and gyrochronology providing the lower bound. Age indicators based on the lithium abundance, X-ray luminosity, kinematics, and isochrone technique are less precise than gyrochronology and chromospheric activity: in addition, for the Li-age estimate it is more difficult to quantify the uncertainty. We therefore do not attempt to combine those with the age range estimated from the gyrochronological and chromospheric activity techniques, although their estimates are consistent with our adopted age of GJ 504. Future asteroseismology studies, or detailed kinematic studies identifying GJ 504 as a member of an MG, may be able to refine our age estimate. The estimated ages are summarized in Table \ref{table: system_property} and Figure \ref{fig: f0_age.eps}.    

\subsection{Other Information}

GJ 504 was previously reported to have a stellar companion at a projected separation of about 650 AU \citep{Poveda1994Binary}, but subsequent observations showed that it does not share a common proper motion; the GJ 504 proper motions are --333.83 $\pm$ 0.25 mas yr$^{-1}$ and 190.24 $\pm$ 0.17 mas yr$^{-1}$ in right ascension and declination \citep{vanLeeuwen2007Hip}, respectively, while those of the stellar companion candidates are 22.9 $\pm$ 11.2 mas yr$^{-1}$ and --51.7 $\pm$ 11.2 mas yr$^{-1}$ in each direction \citep{Roeser2010PPMXL}. Hence, we identify the stellar companion candidate as an unrelated background star. \par

The effective temperature, surface gravity, and metallicity [Fe/H] of GJ 504 were inferred by directly fitting the observed spectrum to the synthetic spectrum \citep{Valenti2005_StellarParam}. Interestingly, GJ 504 has a super-solar metallicity [Fe/H] = 0.28 \citep{Valenti2005_StellarParam}. In order to estimate the mass of the central star, we use an empirical relation of \cite{Torres2010StellarMass}, which is a function of the surface gravity, effective temperature, and metallicity of GJ 504 \citep{Valenti2005_StellarParam}. The error of the estimate was provided by including uncertainties in the spectroscopically determined input parameters, the reported intrinsic scatter in the relation, and correlations of the best-fit coefficients \citep{Fleming2010MARVES}. As a result, the mass of GJ 504 is estimated to be 1.22 $\pm$ 0.08 solar masses. The mass of GJ 504 can also be derived based on matching its spectroscopically determined surface parameters with theoretical stellar evolution models \citep{Takeda2007NSStructure,Valenti2005_StellarParam}. The theoretical models infer the mass of GJ 504 to be 1.29$^{+0.05}_{-0.04}$ \citep{Takeda2007NSStructure} or 1.28 $\pm$ 0.03  $M_{\rm{\odot}}$ \citep{Valenti2005_StellarParam}, overlapping with the empirical estimate. For our analyses, we employ the mass estimate based on the empirical function for GJ 504. \par

Broad band observations of the infrared excess above the photosphere at 24 $\rm{\mu m}$ and 70 $\rm{\mu m}$ place constraints on the distribution of dust in the inner ($<$ 10 AU) and outer disks. A recent Spitzer survey showed that 4\% of Sun-like stars have 24 $\rm{\mu m}$ excess, while 16\% of Sun-like stars have 70 $\rm{\mu m}$ excess \citep{Trilling2008NSDebri}. GJ 504 was shown to not have excesses at either of these band passes ($L_{\rm{dust}}/L_{\rm{star}} < 2.1 \times 10^{-6}$), and therefore does not seem to harbor a significant debris disk \citep{Bryden2006NSDebri}. 

\section{Observations and Data Reduction}  

GJ 504 was observed as part of the Strategic Exploration of Exoplanets and Disks with Subaru (SEEDS) survey \citep{Tamura2009SEEDS}, which aims to detect and characterize giant planets and circumstellar disks using the 8.2-m Subaru Telescope. In 2011 and 2012, we obtained $J$- ($\sim$1.2 $\rm{\mu}$m), $H$- ($\sim$1.6 $\rm{\mu}$m), and $K_{\rm{s}}$- ($\sim$2.1 $\rm{\mu}$m) band images using the High Contrast Instrument for the Subaru Next Generation Adaptive Optics \citep[HiCIAO;][]{Suzuki2010HiCIAO} with AO188, a 188 actuator adaptive-optics system \citep{Hayano2008AO188}. Furthermore, we obtained follow-up $L^{\prime}$-band ($\sim$3.8 $\rm{\mu}$m) images using InfraRed Camera and Spectrograph \citep[IRCS;][]{Kobayashi2000IRCS} and AO188. See Table \ref{table: astrometry} for the summary of GJ 504 observations.

\begin{deluxetable*}{ccccccccccc}
\tabletypesize{\footnotesize}
 \tablecaption{Observation Log and Astrometric Measurements of GJ 504 b.}
 \tablewidth{0pt}
 \tablecolumns{4}
\tablehead{ \colhead{ Obs. Date} & \colhead{Inst.} & \colhead{Band} & \colhead{Total DIT} & \colhead{Mode} & \colhead
{FRA} & \colhead{Proj. Sep.} &\colhead{PA} \\
\colhead {(UT)} & \colhead{} & \colhead{} & \colhead{(min)} & &\colhead{(\arcdeg)} & \colhead{(arcsec)} & \colhead{(\arcdeg)}}
\startdata
   2011 Mar 26 & HiCIAO & $H$ & 12 & ADI & 52.7 & 2.479 $\pm$ 0.016 & 327.94 $\pm$ 0.39 \\
   2011 May 22 & HiCIAO & $H$ & 21 & ADI & 56.9 & 2.483 $\pm$ 0.008 & 327.45 $\pm$ 0.19 \\
   2011 Aug 12 & IRCS & $L^{\prime}$ & 7 & DI & $\cdots$ & 2.481 $\pm$ 0.033 & 326.84 $\pm$ 0.94 \\
   2011 Aug 15 & IRCS & $L^{\prime}$ & 19 & DI & $\cdots$ & 2.448 $\pm$ 0.024 & 325.82 $\pm$ 0.66 \\
   2012 Feb 28	& HiCIAO & $K_{\rm{s}}$ & 38 & ADI & 42.5 & 2.483 $\pm$ 0.015 & 326.46 $\pm$  0.36 \\
   2012 Apr 12	& HiCIAO & $J$ & 40 & ADI & 66.1 & 2.487 $\pm$ 0.008 & 326.54 $\pm$ 0.18 \\
   2012 May 25 	& IRCS & $L^{\prime}$ & 68 & ADI & 129.0 & 2.499 $\pm$ 0.026 &  326.14 $\pm$ 0.61 
     \enddata
      \tablecomments{The dates, used instruments, observation bands, total detector integration times (DIT), observation modes, and field rotation angles (FRA) are summarized for each observation run. The measured projected separations and position angles (PAs) are also described.}
     \label{table: astrometry}
\end{deluxetable*}

\subsection{HiCIAO Observations} \label{sec: HiCIAO observation}

We discovered GJ 504 b in $H$-band observations using HiCIAO on 2011 March 26 with AO188. We used GJ 504, its host, as the natural guide star, and used an atmospheric dispersion corrector (ADC) to prevent it from drifting on the detector due to differential refraction in the visible and near-infrared. We used a circular occulting mask 0\farcs4 in diameter and angular differential imaging \citep[ADI;][]{Marois2006ADI} to remove the starlight and the stellar speckles. ADI keeps the telescope pupil fixed relative to the camera, while the field of view (FoV) rotates as the target moves across the sky. As a result, diffraction and speckle patterns in the stellar point spread function (PSF), caused by the major part of telescope optics, remain fixed on the detector while real on-sky sources, such as a planet, rotate about the natural guide star. This allows ADI data processing to distinguish between stellar speckles and real objects, and to remove the starlight while preserving potential planets. \par       

We conducted several follow-up observations in 2011 and 2012 using the same instrument and configuration. $H$-band follow-up observations were performed on 2011 May 22. The new $H$-band observations had both a longer total exposure time and better AO performance than the March observations, giving a higher signal-to-noise ratio ($\rm{S/N}$) in the final images. We obtained the $K_{\rm{s}}$- and $J$-band data in 2012 February and April. The seeing in the $J$-band observation was among the best ever measured during SEEDS observations, and the AO worked very well, but those in $K_{\rm{s}}$-band observations were worse than $J$- and $H$-band observations. \par

To correct the optical distortions of HiCIAO (and IRCS), we observed the globular cluster M5 using both instruments and compared the results to distortion-corrected images taken using the Advanced Camera for Surveys (ACS) on the {\it Hubble Space Telescope}; the accuracy and calibration for plate scale, orientation angle, and distortion of ACS/WFC camera, whose data were used for our calibration, have been previously reported \citep{ACSPlateScale2007}. The distortion-corrected plate scale was 9.500 $\pm$ 0.005 mas pixel$^{-1}$ for HiCIAO, and its north orientation was 0\fdg35 $\pm$ 0\fdg02. Uncertainties in the distortion correction are less than 4 mas for HiCIAO within the central 10$\arcsec$ $\times$ 10$\arcsec$ of the FoV. We have included all of these uncertainties in our positional measurements for the planet detected in both HiCIAO and IRCS images, though they have little effect on our results. Photometric calibrations were performed by observing the central star itself with neutral density (ND) filters with transmissions of 0.590, 0.0628, and 1.14\% for $J$-, $H$-, and $K_{\rm{s}}$-bands, respectively. 

\subsection{IRCS Observations}

To obtain data at a longer wavelength, we observed GJ 504 in the $L^{\prime}$-band using IRCS on 2011 August 12 and 15. We did not use either ADI or an occulting mask during these long-wavelength observations. In addition, we obtained each frame by dithering the target within the FoV, which allowed us to better subtract the high thermal background (see below). In order to confirm whether we can obtain advantages for $L^{\prime}$-band observational results by employing ADI, we performed additional $L^{\prime}$-band observations in 2012 May 25. The longer total exposure time, compared to the 2011 observations, as well as the usage of ADI, led to a higher $\rm{S/N}$. \par

As in the case of HiCIAO, we observed M5 to obtain the calibration data for correcting the distortion of IRCS. After the distortion-correction, IRCS has the plate scale of 20.54 $\pm$ 0.03 max pixel$^{-1}$, and its north orientation was tilted by 0\fdg28 $\pm$ 0\fdg09. Uncertainties in the distortion correction are less than 8 mas for IRCS within the central 10$\arcsec$ $\times$ 10$\arcsec$ of the FoV. Because the $L^{\prime}$-band magnitude of GJ 504 has not been measured, we calibrated it using a standard star observed right before GJ 504 observations. As with HiCIAO observations, we then measured the flux of GJ 504 with an ND filter with a transmission of 0.67\%.

\subsection{High-contrast Data Reduction}

\subsubsection{HiCIAO Data Reduction}

For the HiCIAO data reduction, we first removed stripe patterns \citep[see][]{Suzuki2010HiCIAO} emerged on each data frame, and subsequently performed the flat-fielding. After correcting the image distortion, we registered each dataset to the centroid of GJ 504 A. For the HiCIAO observations, which used an opaque 0\farcs4 diameter occulting mask, we estimated the relative frame centers by calculating cross-correlations of the frames as a function of positional offset. The AO188 and ADC kept the PSF extremely stable and well centered. For confirmation, we fit linear and quadratic functions to the frame-to-frame positional offsets and estimated the positional drifts that remained after AO/ADC correction. The systematic drifts were less than 10 mas for all observations except $K_{\rm{s}}$-band (17 mas), with residual rms scatters of $\sim$2--10 mas. These random positional jitters can be attributed to our registration inaccuracies. \par

We calibrated the absolute centers using a sequence of unsaturated, unmasked exposures taken with an ND filter. For the $H$-band data, the positional reference was acquired right after the end of the masked sequence. In the cases of the $J$- and $K_{\rm{s}}$-band observations, we obtained the positional calibration data in a similar way as for the $H$-band observations, but with the calibration observations interspersed through the science sequence. The positional errors for the unsaturated data ranged from $\sim$2 to $\sim$5 mas, where the lower quality PSFs for the $H$-band data obtained in 2011 March observations or the $K_{\rm{s}}$-band data in 2012 February caused the largest errors. The unsaturated data used a combination of ND filters and field lens that were different from the main science data, possibly introducing a small systematic offset in position. To measure offsets introduced by the insertion of ND filters, we obtained images of a pinhole mask at the focal plane of HiCIAO, with and without ND filters, and measured the offsets. To determine offsets introduced by the use of a different field lens, we examined stars in images of the M5 globular cluster, observed both with and without the mask field lens. We measured these offsets to be small, $<$ $\sim$3 mas. The values of the offsets introduced by the mask field lens were confirmed with three binaries in the SEEDS samples. We did not attempt to remove the offsets, but we have instead considered both of the offsets in the measurements of positions by adding systematic errors of 3 mas to the image registration uncertainties. In total, for $H$-band observations, image registration conservatively contributed $\sim$12 mas to our reported positional uncertainties for GJ 504 b in March and $\sim$6 mas in May. Those uncertainties are $\sim$10 and $\sim$6 mas in 2012 February ($K_{\rm{s}}$) and April ($J$), respectively. \par

We used the Locally Optimized Combination of Images (LOCI) algorithm \citep{Lafreniere2007LOCI} to reduce the HiCIAO data. LOCI constructs a model PSF for each frame using the other frames in that dataset as references. The resulting PSF is locally a linear combination of the reference images, with the coefficients determined using a least-squares minimization. To avoid subtracting astrophysical sources, LOCI only uses reference frames with a minimum level of field rotation. For our reductions of GJ 504, we required at least 1.2 (=$N_{\rm{\sigma}}$) $\times$ FWHM / $R$ ($\times$ 180/$\pi$ deg) of field rotation, where the FWHM of PSFs on HiCIAO images are $\sim$50--60 mas and $R$ is the radial distance from GJ 504. As a result, any physical source was displaced by at least 1.2 FWHM between the working frame and the reference frames. We investigated if the $\rm{S/N}$ was improved by applying $N_{\rm{\sigma}}$ = 1.5 to the reductions but the $\rm{S/Ns}$ remained almost the same or were slightly reduced compared with the cases for $N_{\rm{\sigma}}$ = 1.2, and we thus carry out all analyses based on the images reduced with $N_{\rm{\sigma}}$= 1.2.\par

For the $H$-band observations, our analysis of the data from March 26 detected a faint companion at an $\rm{S/Ns}$ of 5, where we calculated the noise as the standard deviation of the intensity in concentric annuli about GJ 504. Because of the longer integration time and better observing conditions for the observations in 2011 May, we detected GJ 504 b with $\rm{S/Ns}$ = 9.0. We detected no other companion candidates ($\rm{S/Ns}$ $\leq$ 5) in either observation. GJ 504 b is clearly confirmed with $\rm{S/Ns}$ of 18.4 at $J$-band, while the $\rm{S/Ns}$ of the $K_{\mathrm{s}}$-band observations is 4.6.

\begin{figure*}
	\plotone{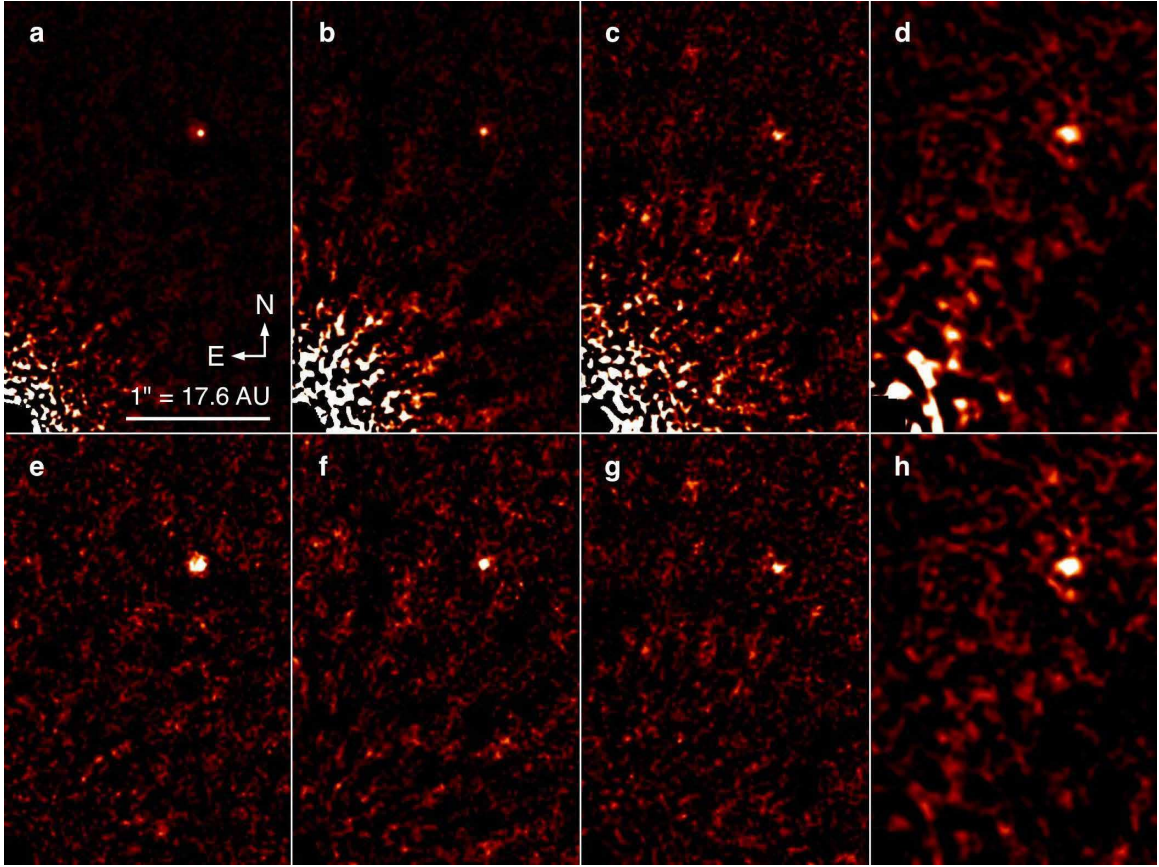}
	\epsscale{0.8}
\caption{ $J$ ($\sim$1.2 $\rm{\mu}$m), $H$ ($\sim$1.6 $\rm{\mu}$m), $K_{\rm{s}}$ ($\sim$2.2 $\rm{\mu}$m), and $L^{\prime}$-band ($\sim$ 3.8 $\rm{\mu}$m) images of the newly discovered exoplanet, GJ 504 b. The four top panels show images reduced with the LOCI pipeline (\textbf{a}: $J$, \textbf{b}: $H$, \textbf{c}: $K_{\rm{s}}$, \textbf{d}: $L^{\prime}$). The corresponding signal-to-noise maps are shown on the four bottom panels (\textbf{e}: $J$, \textbf{f}: $H$, \textbf{g}: $K_{\rm{s}}$, \textbf{h}: $L^{\prime}$), in which the planet is detected with signal-to-noise ratios of 18.4, 9.0, 4.6, and 8.0, respectively. All signal-to-noise maps are shown at a stretch of [--5, 5]. In all panels, the star is located approximately in the lower left corner. North is up and east is left. }
\label{fig: f1.eps}
\end{figure*}

\begin{figure*}
 \plotone{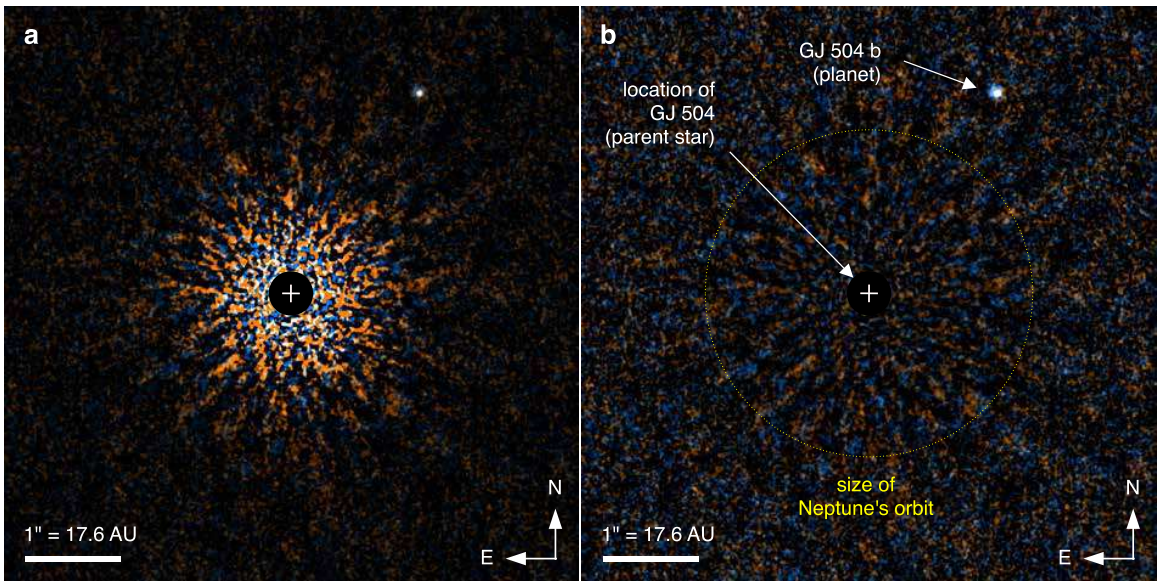}
\epsscale{6}
\caption{Discovery images of the exoplanet GJ 504 b. The central 6$\arcsec$ $\times$ 6$\arcsec$ of two different high-contrast images were overlaid for this false color-composite image: orange represents $H$-band ($\sim$1.6 $\rm{\mu}$m; Subaru/HiCIAO; 2011 May 22), and blue represents $J$-band ($\sim$1.2 $\rm{\mu}$m; Subaru/HiCIAO; 2012 April 12). The $J$-band image was rotated by 0\fdg9 to compensate for the planet's observed orbital motion (cf. Figure \ref{fig: f4_astrometry.eps}). Panel \textbf{a} shows the intensity after suppressing flux from the central star. Panel \textbf{b} shows the associated signal-to-noise ratio ($\sim$9 in $H$; $\sim$18 in $J$). The orbital radius of Neptune ($\sim$30 AU) is shown for comparison with our own solar system. The planet GJ 504 b is clearly visible as a white spot at a projected distance of 43.5 AU from the star GJ 504. The white color implies that the planet signal is persistent in both observations, setting it apart from the uncorrelated residual noise in each of the constituent images. }
\label{fig: f2.eps}
\end{figure*}

The $J$, $H$, and $K_{\rm{s}}$-band images for GJ 504 b are shown in Figure \ref{fig: f1.eps}. Also, the high-contrast two-color composite images is shown in Figure \ref{fig: f2.eps}, in which GJ 504 b is clearly visible 2\farcs48 northwest of GJ 504. \par

\subsubsection{IRCS Data Reduction}

For the IRCS data, we first corrected flat-fielding. Subsequently, we subtracted the thermal background and then used an algorithm \citep{Galicher2011HR8799} similar to LOCI in order to further reduce the background. We estimated the background value at each pixel and in each frame by dithering the images and using a linear combination of the frames in which the central source, GJ 504, is sufficiently displaced. As in LOCI, we computed the coefficients for this linear combination using a least-squares fit. Eventually, we subtracted the resulting estimate of the local background whose variation is a function of time. As in HiCIAO reductions, we subsequently corrected the distortion of IRCS images. \par

The IRCS $L^{\prime}$-band data were taken without a mask, and were saturated out to a radius of about 3--4 pixels ($\sim$60--80 mas, 0.6--0.8 FWHM, where FWHM is the full width at half maximum of the PSF). We first performed relative registration by cross-correlating these saturated images with one another. We then centroided a set of unsaturated reference images, and cross-correlated the mean of these unsaturated PSFs with each saturated frame to measure the positional offset. Finally, we applied this average offset to each of the saturated frames. The rms scatter of the offsets was $\sim$6 mas for both observations in 2011 August, and $\sim$4 mas for 2012 May. We adopt these scatters as the registration uncertainties in $L^{\prime}$-band data. \par

We then combined these registered frames. We detected GJ 504 b with an $\rm{S/Ns}$ of 3 and 4.3 in the August 12 and August 15 data, respectively. Because the data for 2012 May 25 were obtained with ADI, we could use LOCI for the PSF subtraction before stacking the registered data: the LOCI reduction parameters were set to be the same as the case of HiCIAO. In the resulting image, the companion was clearly detected with an $\rm{S/Ns}$ of 8.0 (see Figure \ref{fig: f1.eps}).  

\begin{figure*}[thp]
	\plotone{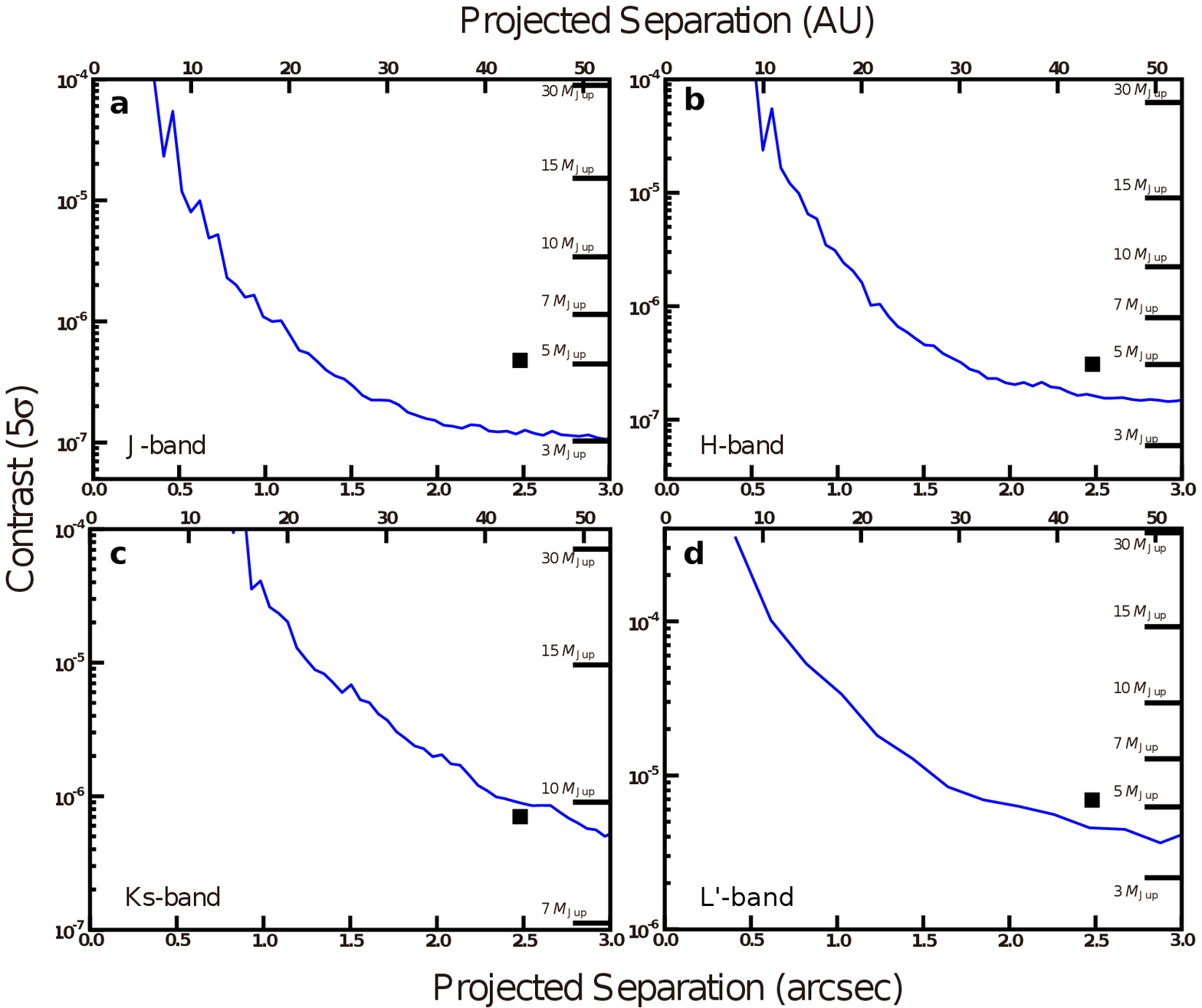}
\caption{Detection limits for $J$, $H$, $K_{\rm{s}}$, and $L^{\prime}$-bands. Each panel shows 5$\sigma$ detection limits (blue curves) estimated with reduced $J$, $H$, $K_{\rm{s}}$, and $L^{\prime}$-band data. The left vertical axes indicate measured contrasts relative to the primary at each separation. The corresponding detectable mass limits, at the system age of 300 Myr using COND models \citep{Baraffe2003COND}, are shown along the right axes. The bottom horizontal axes show angular separations projected from the primary in arcseconds. The top axis gives separations in AU. The contrasts for GJ 504 b are indicated by black squares in each panel. \textbf{a}, $J$-band detection under exceptionally good weather and seeing conditions; \textbf{b}, $H$-band; \textbf{c}, $K_{\rm{s}}$-band; \textbf{d}, $L^{\prime}$-band. }
\label{fig: f3_contrast.eps}
\end{figure*}

\subsection{Positional Measurements of GJ 504 b}

We measured the position of GJ 504 b by fitting two-dimensional elliptical Gaussian functions to the final reduced data. Uncertainties in these Gaussian centroids range from 2 to 44 mas, and dominate the total uncertainties in the position of GJ 504 b. Sometimes, LOCI introduces artificial shifts of the companion positions \citep{Soummer2011HSTHR8799}. Measuring positions of artificial companions, which have the similar brightness and locations with the GJ 504 b, we have investigated how large offsets are added to the positions of GJ 504 b. Because our adaptation for LOCI parameters is not aggressive and the planet's separation is sufficiently large from the star ($r$ = 2\farcs48), the astrometric biases are not large.\footnote[6]{Typically, averages of amplitudes for offsets are smaller than 3 mas for all epochs of data, and the rms scatters are also sufficiently small compared to our assigned total positional errors. We do not add the possible systematic errors due to LOCI processing to our astrometric total errors since our measurements for the systematic offsets are degenerate to other error sources such as speckles or photon noise. However, if we include those in final astrometric errors, they have no impact on our conclusions.} The positional measurements are converted to the projected separations and position angles (P.A.s) relative to the central star, and all of those are summarized in Table \ref{table: astrometry}. Then, the uncertainties of measured separations and P.A.s were calculated by including the possible individual error sources of the image registration, distortion correction, plate scale and orientation angle of each camera, and PSF-fit for GJ 504 b.     

\subsection{Photometry} \label{sec: photometry}

We performed aperture photometry on all of our observations of GJ 504 b. For HiCIAO observations of the star GJ 504, we adopted magnitudes of 4.11, 3.86, and 3.81 for its $J$-, $H$-, and $K_{\rm{s}}$-band brightness \citep{Kidger2003}. The $L^{\prime}$-band brightness of GJ 504 was calibrated with a standard star, so that we obtained $L^{\prime}$ = 3.94 $\pm$ 0.09 mag for GJ 504. For all photometries of both the primary and the planet, we set the aperture radius to be 1.0 PSF FWHM, which ranged from 47 mas in $J$ to 100 mas in $L^{\prime}$.  

By subtracting a (local) linear combination of frames from each image, LOCI can artificially reduce the flux of a point source \citep{Lafreniere2007LOCI}. We limit this effect by choosing a relatively large $N_{\rm{\sigma}}$ and estimate its magnitude using artificial point sources. We inject point sources in the HiCIAO data with projected separations and fluxes similar to those of GJ 504 b, measuring their magnitudes before and after applying LOCI. We measured flux losses ranging from 13\% to 23\% for all LOCI reductions, where the typical scatter for the measurements of flux losses was 5\%. We added this value to the final photometric errors. To verify this result, we re-reduced the HiCIAO data by subtracting a median PSF \citep{Marois2006ADI}, but not applying LOCI. The resulting fluxes for GJ 504 b had larger errors, but were consistent with their values in the final, LOCI-processed data. We estimated all of our photometric errors for the companion by calculating the rms scatter of aperture fluxes in blank-sky regions at the separation of GJ 504 b. For the $H$-band data, these photometric errors were larger than the flux difference between the two sets of observations. For $H$-band observations, the unsaturated frames used to calibrate the companion fluxes were obtained before and after the main science sequences with the mask, while the unsaturated integrations for $J$-, $K_{\rm{s}}$-, and $L^{\prime}$-band were conducted in intervals among the main sequences as well. The total number of unsaturated frames that were used for photometric calibration was more than 31 for each epoch observation, except in 2011 March. We measured standard deviations of photometry of the primary using the unsaturated data; the worst value for the standard deviation was 22\% in 2011 March, while these values were less than 12\% in all other observations. We include these variations in the final photometric uncertainties, in addition to the errors of photometry for companion and self-subtraction estimates. For $L^{\prime}$-photometry, we had three sets of images, but did not combine the result of 2012 with those derived from the data obtained in 2011, because the $\rm{S/Ns}$ of the $L^{\prime}$ images in 2011 were significantly worse than for the 2012 images; in the former case, no ADI was used and the observing conditions were very poor, with airmasses as high as 3.5 at maximum. Therefore, our adopted value for the $L^{\prime}$ photometry is based exclusively on the data obtained in 2012. The photometric measurements of GJ 504 b at each band are summarized in Table \ref{table: system_property}, with the $H$-band flux computed using the weighted mean of the March and May values. Furthermore, the detection limits estimated from the reduced data are shown in Figure \ref{fig: f3_contrast.eps}. 

\section{Results}

\subsection{Astrometry}

\begin{figure}
\includegraphics[width=1.02 \linewidth]{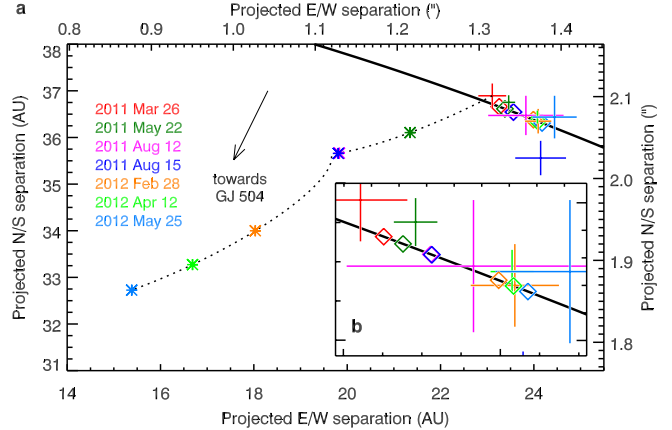}
\caption{Astrometric analysis of GJ 504 b. The measured positions of the planet relative to its parent star at the seven epochs of observation are shown as plus signs (size corresponds to error bars). A background star would follow the dotted trajectory, with star symbols marking the position at the epochs of observation. More information on the orbital simulation can be found in the Appendix. Diamond symbols mark the position of the planet along the most likely orbit at the epochs of observation. Plus, star, and diamond symbols are color-coded for epoch. \textbf{a}, View of the planet's motion compared with the track expected if GJ 504 b were a background star; \textbf{b}, zoomed-in view of the observed positions of GJ 504 b.}
\label{fig: f4_astrometry.eps}
\end{figure}

To confirm that GJ 504 b is gravitationally bound to the host star, we tracked its proper motion over the course of more than a year, from 2011 March through 2012 May (Figure \ref{fig: f4_astrometry.eps}). Figure \ref{fig: f4_astrometry.eps} allows to clearly visually distinguish the measured positions from the track expected for a background object. However, we also carried out a $\rm{\chi^2}$ test to derive a statistical  confidence that the source is a background star. This yielded a value of $\chi^2 = 1,340$ (12 dof). Therefore, the $\rm{\chi^2}$ test on the astrometry decisively excludes GJ 504 b as an unrelated background source. On the other hand, if we assume GJ 504 b to be an interloping foreground object, its color ($H-L^{\prime}$ = 3.3 mag) would rule out a star anyway, leaving only the possibility of a cool brown dwarf ($T < 600$ K) within a distance of $\sim$30 pc. Even under optimistic assumptions about the brown dwarf mass function \citep{Kirkpatrick2011WISEBD}, less than one such object exists per 10 deg$^{2}$, and the probability of a chance alignment within 3\arcsec\ of GJ 504 is no more than 10$^{-7}$. \par

Furthermore, in order to demonstrate that the measured motion of GJ 504 b relative to its parent star GJ 504 is indeed consistent with the expected orbital motion of a gravitationally bound planet, we have fitted a Keplerian curve to its measured positions relative to the star, and constrained its orbit (see Appendix). Here we assumed that the star has 1.22 $M_{\rm{\odot}}$. We show the best-fit curve in Figure \ref{fig: f4_astrometry.eps}. The best-fit $\rm{\chi^2}$ was derived to be 11.7 (8 dog), which confirms that the measured positions relative to the central star are consistent with dynamical orbital motion around the star.\footnote[7]{The position measured in 2011 August 15 deviates from the best-fit orbit, which increases the $\rm{\chi^2}$. That measurement was obtained without ADI at high airmass.} These analyses robustly prove that GJ 504 and GJ 504 b comprise a physically associated system. 

\subsection{Mass Estimate} \label{section: mass estimate}

\begin{figure}
\plotone{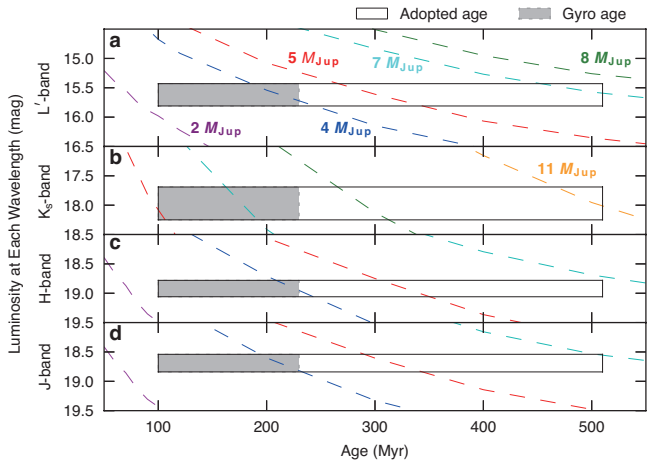}
\caption{Mass of GJ 504 b, as derived from the theoretical relation between the planetary age and luminosities for various wavelengths. Estimated age (Myr) and near-infrared luminosities (absolute magnitudes) for GJ 504 b are plotted over cooling curves for planets of various masses calculated from the hot-start model \citep{Baraffe2003COND}; \textbf{a}, $L^{\prime}$-band ($\sim$3.8 $\rm{\mu}$m) luminosity; \textbf{b}, $K_{\rm{s}}$-band ($\sim$2.1 $\rm{\mu}$m); \textbf{c}, $H$-band ($\sim$1.6 $\rm{\mu}$m); \textbf{d}, $J$-band ($\sim$1.2 $\rm{\mu}$m). The vertical size of each box shows the estimated luminosity range ($\sim$ 1$\sigma$) in each band.  The gyrochronological age (indicated by the gray box) is the most direct age estimate among all age indicators available for GJ 504; it implies a planet mass of 4$^{+1.5}_{-1.0}$ $M_{\rm{Jup}}$. The open box with the solid black line boundary indicates the wider age range encompassing all of the adopted age estimates, and implies a mass of 4$^{+4.5}_{-1.0}$ $M_{\rm{Jup}}$.}
\label{fig: f5_hotstart.eps}
\end{figure}

Using the luminosity and age of GJ 504 b, we thus proceed to estimate its mass based on the hot-start model \citep{Baraffe2003COND} (see Figure \ref{fig: f5_hotstart.eps}). We then calculate a mass by comparing luminosity of GJ 504 b at each band to the models, and eventually average the mass estimates at $J$, $H$, $K_{\rm{s}}$, and $L^{\rm{\prime}}$-bands, which provides a mass of the planet, and its error range, which includes the errors from age and photometry. Any bias in the mass estimate caused by focusing on the photometry in a particular band should be mitigated when taking the average of the photometry of all bands. For the upper bound, we estimate the mass of GJ 504 b from the 1$\sigma$ bright end of the photometry at each band, using hot-start models for an age of 510 Myr. The collective upper limit is then acquired by averaging the estimates at each band. The lower limit is calculated in an equivalent way, using the 1$\sigma$ faint end photometry and an age of 100 Myr. Likewise, the best estimate uses the mean photometry in each band and an age of 160 Myr. As a result of this procedure, a mass estimate of 4$^{+4.5}_{-1.0}$ $M_{\rm{Jup}}$ is derived. The age uncertainty dominates the errors. If we adopt the gyrochronological age of 160$^{+70}_{-60}$ Myr, the most direct of our age indicators, the mass of GJ 504 b is constrained to lie between 3 and 5.5 $M_{\rm{Jup}}$. This is well below 13.6 $M_{\rm{Jup}}$, the deuterium-burning threshold commonly used to divide planets and brown dwarfs \citep{Marois2008HR8799, Burrows1997} and among the lowest masses for exoplanets discovered by direct imaging.\par

Adopting the $K_{\rm{s}}$-band luminosity may result in a bias in the mass estimation, since this wavelength range is particularly sensitive to atmospheric parameters such as surface gravity and metallicity \citep{Fortney2008hotstart, Spiegel2012ColdStart, Janson2011GJ758B}. Indeed, \citet{Fortney2008hotstart} predict that raising the metallicity to five times solar metallicity results in a $K_{\rm{s}}$-band brightening of a factor of 2--6 for a planet with temperature and surface gravity values in the range of what is expected for GJ 504 b. Since the star GJ 504 has a metallicity above the solar value, it is very likely that the planet's metallicity is also enhanced. For reference, if we neglect the $K_{\rm{s}}$-band photometry, the best-fit mass of GJ 504 b becomes 3.5 $M_{\rm{Jup}}$, and is confined to be less than 8 $M_{\rm{Jup}}$ based on the adopted age-range. \par

\begin{figure*}
\epsscale{0.9}
\plotone{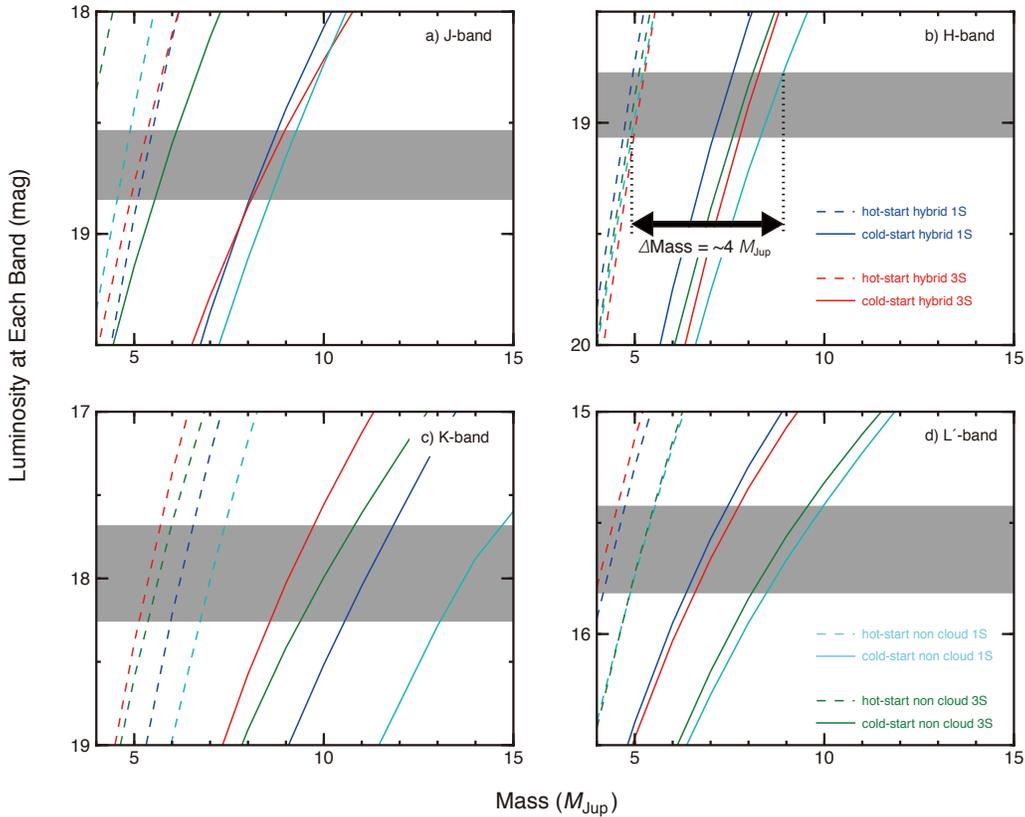}
\caption{Differences between mass estimates from cold-start models and hot-start models for 100 Myr old planets. The mass-luminosity relations modeled in \citet{Spiegel2012ColdStart} are illustrated with dashed and solid lines for hot-start (entropy $=13\ k_{\rm{B}}$ baryon$^{-1}$) and cold-start (8 $k_{\rm{B}}$ baryon$^{-1}$) initial conditions, respectively. We show the relations for four atmosphere types on each panel (\textbf{a}: $J$, \textbf{b}: $H$, \textbf{c}: $K_{\rm{s}}$, \textbf{d}: $L^{\prime}$): ``hybrid'' means patchy cloud model and 1 or 3 S means one or three times solar-abundance model. The gray regions correspond to the absolute magnitudes of GJ 504 b in each band and their error ranges ($\pm 1\sigma$). The lower age limit (100~Myr) of GJ~504 is chosen in this illustration since that is the stage at which the largest differences occur between hot-start and cold-start conditions. As an example, the arrow in panel \textbf{b} indicates the difference in mass arising from different initial conditions when the 1S cloud-free model atmosphere is adopted. The difference is $\sim$4 $M_{\rm{Jup}}$, which is also the median result for the full set of atmospheres and photometric bands.} 
\label{fig: hot_cold_comparison_v1.eps}
\end{figure*} 

\begin{figure*}
\plotone{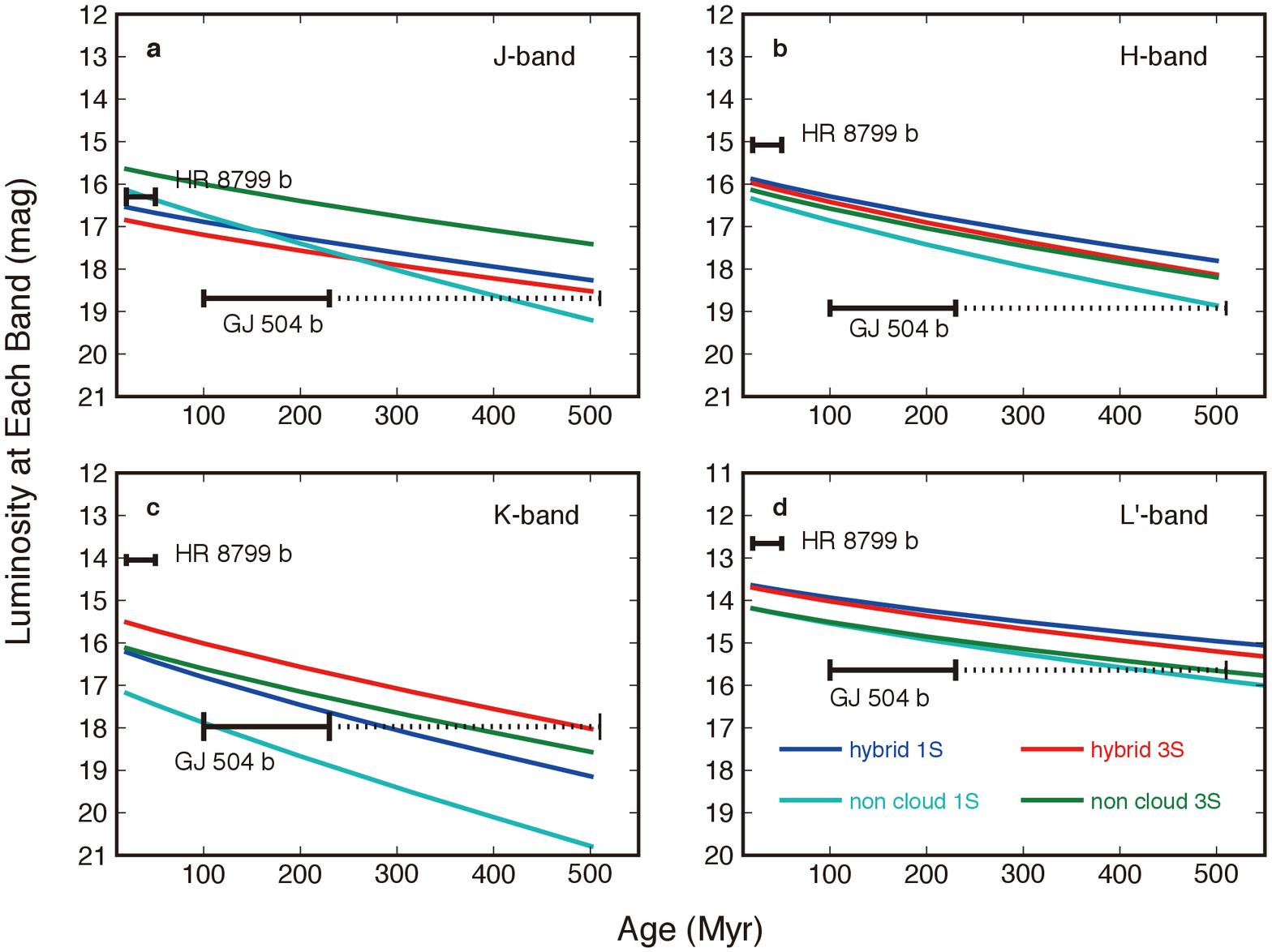}
\caption{Comparison of multiple wavelength magnitudes of GJ 504 b with cold-start model \citep{Spiegel2012ColdStart} predictions at the planet/brown dwarf boundary. In each panel, the cooling tracks predicted by the coldest-start models for a 14 $M_{\rm{Jup}}$ planet are plotted in the age range of 20--500 Myr. Different panels correspond to the different bands (\textbf{a}: $J$, \textbf{b}: $H$, \textbf{c}: $K_{\rm{s}}$, \textbf{d}: $L^{\prime}$). In addition, we show the cooling curves for four atmosphere types, where ``hybrid'' means patchy cloud model and 1 or 3 S means one or three times solar-abundance model \citep{Spiegel2012ColdStart}. The age range of GJ 504 b is indicated by solid and dotted lines. The solid lines and dotted lines correspond to the age range inferred with gyrochronology (160$^{+70}_{-60}$ Myr) or chromospheric activity (330 $\pm$ 180 Myr), respectively. For comparison, we plot the brightness and age of HR 8799 b (30$^{+20}_{-10}$ Myr) \citep{Marois2008HR8799, Metchev2009HR8799} in each panel.}
\label{fig: f6_cold14Mjup.eps}
\end{figure*}

The robustness of mass estimates for directly imaged exoplanets is usually limited by theoretical models used to derive age-luminosity relations. One uncertainty in these relations is the choice of initial conditions, which influences the estimation of a planetary mass. Since GJ 504 is older than 100 Myr, its inferred mass is less sensitive to the uncertainty in initial conditions for the employed model \citep{Marley2007hotstart, Spiegel2012ColdStart} than the previously imaged planets which have younger age than GJ 504 b. To test the sensitivity of the GJ 504 b mass to uncertainties in the hot-start model, we compare the near-infrared luminosities of GJ 504 b with a suite of models with ages between 100 and 510 Myr, initial thermal states ranging from hot to extremely cold (initial specific entropies from 8.0 to 12--13 $k_{\rm{B}}$ baryon$^{-1}$), masses between 1 and 20 $M_{\rm{Jup}}$, varying cloud properties, and metallicities between one and three times the solar value \citep{Spiegel2012ColdStart}. Even when adopting an age of 100 Myr, for which the largest differences arise between the hottest- and coldest-start models, the difference in the mass estimates between the two cases is typically $\sim$4 $M_{\rm Jup}$, which is calculated as the median of all estimates for each different model atmosphere and each band (see Figure \ref{fig: hot_cold_comparison_v1.eps}). At older ages, the difference decreases even further. By contrast, if we use HR 8799 b \citep[20--50 Myr;][]{Marois2008HR8799, Marois2010HR8799} as an example of planets younger than 50 Myr, the corresponding difference is higher than 10 $M_{\rm{Jup}}$, which is significantly larger than the case of GJ 504 b. Thus, in quantitative terms, the dependency on initial conditions for GJ 504 b is indeed rather week. This also allows to confine the mass of GJ 504 b to within the planetary mass regime, essentially regardless of initial conditions (see Figure \ref{fig: f6_cold14Mjup.eps}). The mass inferred from each of the $J$-, $H$-, and $L^{\prime}$-bands individually is practically always below the deuterium burning limit of $\sim$14 $M_{\rm{Jup}}$, apart from the most extreme cases in which both the coldest initial conditions and ages near the oldest end of the age range are adopted simultaneously. Note that these extreme cold-start models may be unrealistic for giant planets \citep{Spiegel2012ColdStart}. 

\section{Discussion}

Here, we discuss the unique properties of GJ 504 b through comparison with other imaged planets. Furthermore, the possible origin of GJ 504 is discussed.  

\subsection{Previously Imaged Planet Candidates}

Some of the previously imaged exoplanets were discovered around young (10--50 Myr), relatively massive stars like HR 8799 \citep[1.5 $M_{\rm{\odot}}$; 20--50 Myr;][]{Marois2008HR8799, Marois2010HR8799} and $\beta$ Pictoris \citep[1.75 $M_{\rm{\odot}}$; 10--20 Myr;][]{Lagrange2010BetaPic}, with masses estimated at 5--9 $M_{\rm{Jup}}$, assuming hot-start conditions. Recently, SEEDS detected a $\sim$13 $M_{\rm{Jup}}$ companion around the B9-type star Kappa And \citep[2.5 $M_{\rm{\odot}}$; 20--50 Myr;][]{Carson2012}. An exoplanet candidate around another massive star, Fomalhaut \citep[2.0 $M_{\rm{\odot}}$; $\sim$200 Myr;][]{Kalas2008Formauhalt}, has been detected at optical wavelengths. The images were reanalyzed by \citet{Galischer2013_Fomalhaut} and \citet{Currie2012_Fomalhaut}, who recovered the point source at an additional wavelength. In addition, new optical images were recently presented with a report of an eccentric orbit \citep{Kalas2013Fomalhaut}. Very recently, another imaged planet candidate has been reported around the young star HD 95086 (10--17 Myr; $\sim$1.6 $M_{\rm{\odot}}$), but a robust second epoch confirmation still needs to be acquired \citep{Rameau2013HD95086}. Other giant planet candidates have also been imaged at wide separations around very young ($<$10 Myr) stars \citep[see e.g.,][]{Kuzuhara2011SR12C}. Among these, 1RXS J1609 b \citep{Lafreniere20081RXJ1609} orbits a pre-main sequence solar-mass star at a wide separation of 330 AU. \par

For all the aforementioned companions, their masses have been inferred from comparisons of their infrared luminosity with hot-start models. However, due to the youth of those systems, they are highly affected by hot- versus cold-start conditions. The cold-start models imply that even HR 8799 b, which may have the lowest masses among them, would be more massive than $\sim$14 $M_{\rm{Jup}}$ (See Figure \ref{fig: f6_cold14Mjup.eps}): its mass is estimated to be $\sim$20 $M_{\rm{Jup}}$ or higher (except at $J$-band). Therefore, under cold-start assumptions, all the exoplanets around these young stars have much higher masses than currently inferred, and are pushed into the brown dwarf mass regime ($> \sim$14 $M_{\rm{Jup}}$). One special case is the candidate around Fomalhaut, for which a mass below 3 $M_{\rm{Jup}}$ has been reported, but deep infrared non-detections indicate that the detected light from the candidate cannot arise from a planetary surface \citep{Janson2012Folmauhaut}; hence, its physical nature remains unclear. Regardless of the caution for the hot-start models, a few independent constraints, such as dynamical stability analysis in the case of the HR 8799 system \citep[e.g.,][]{Fabrycky2010_HR8799,Marois2010HR8799,Sudol2012_HR8799}, suggest that the hot-start models may be realistic. \par

As discussed above, the mass estimates for directly-imaged planets suffer from uncertainties of cooling models for giant planets until future calibrations may be obtained by comparing model-dependent masses with improved dynamical mass estimates. However, due to its old age, GJ 504 b is less dependent on the models' uncertainty related to the hot versus cold conditions, compared with previously imaged planets. 

\subsection{Color of GJ 504 b}

\begin{figure*}
\plotone{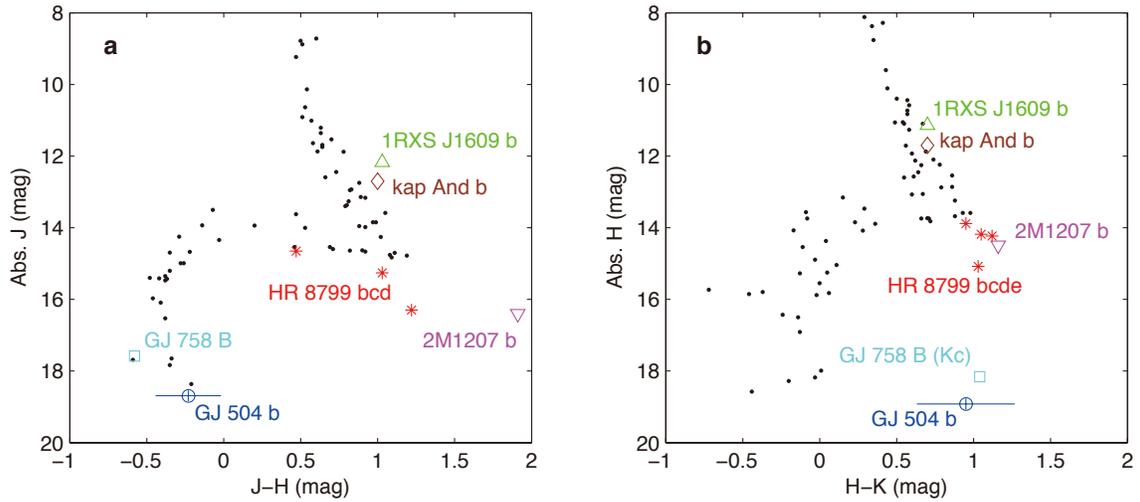}
\caption{Color-magnitude diagrams for GJ 504 b. Shown for comparison are LT-type field brown dwarfs as black dots \citep{Leggett2010LTdwarf}, and a sample of planet and brown dwarf companions from the literature. Panel \textbf{a} shows $J$-band luminosity ($J$-band absolute magnitude) vs. $J-H$ and panel \textbf{b} shows $H$-band luminosity vs. $H-K_{\rm{s}}$ ($H-K_{c}$ for GJ 758 B). Directly imaged planets are each labeled.}   
\label{fig: f7_CMD.eps}
\end{figure*} 

We can compare the magnitudes and colors of GJ 504 b to similar objects in the literature in order to get a sense of the physical properties of this planet. For this purpose, it has been placed on color--magnitude diagrams (Figure \ref{fig: f7_CMD.eps}) along with LT-type field brown dwarfs \citep{Leggett2010LTdwarf} and HR 8799 bcde \citep{Marois2010HR8799, Metchev2009HR8799, Skemer2012HR8799}, Kappa And b \citep{Carson2012}, 1RXS J1609 b \citep{Lafreniere20081RXJ1609}, 2M 1207 b \citep{Chauvin20042M1207, Mohanty20072M1207, Skemer20112M1207}, and GJ 758 B \citep{Thalmann2009GJ758B, Janson2011GJ758B}. 

Given the location of GJ 504 b in the color-magnitude diagrams, it is clear that it is much colder than previous planet detections. Indeed, an application of $JHK_{\rm{s}}L^{\prime}$ luminosities of GJ 504 b to those predicted by the cooling models of \citet{Baraffe2003COND} implies an effective temperature of 510$^{+30}_{-20}$ K for the planet, where we estimated the best estimate and lower/upper limits in the same way as for the mass estimate of GJ 504 b (see Section \ref{section: mass estimate}). The inferred effective temperature is significantly colder than the values of 800--1800 K inferred for previously imaged planets \citep{Marois2008HR8799, Chauvin20042M1207, Bonnefoy2011BetaPic}. The extrasolar giant planets other than GJ 504 are primarily located around the so-called LT transition, which corresponds to the transition from cloudy (L-type) to clear (T-type) atmospheres. GJ 504 b is consistent with having a largely clear atmosphere analogous to the bottom of the T-type sequence, or the top of the yet colder Y-type sequence. It deviates from this sequence in $H-K_{\rm{s}}$, due to overluminosity in $K_{\rm{s}}$-band, which is probably due to a low surface gravity or enrichment in heavy elements, both of which can reproduce such an effect \citep{Burrows2006LTtransition}. These factors have been shown to possibly explain the same trend seen in GJ 758 B \citep{Janson2011GJ758B}, which is an older and more massive companion to a Sun-like star. We note that GJ 504 b was observed with the $K_{\rm{s}}$-band filter and GJ 758 B with the $K_{\rm{c}}$-band filter, which prevents a one-to-one comparison. However, models incorporating heavy element enrichment appear to be a promising option for reproducing the full set of magnitudes of GJ 504 b.

It has been proposed that planetary candidates with hot atmospheres ($T_{\rm{eff}}$ $\sim1000$ K) around HR 8799 or 2M 1207 may have thick clouds \citep{Barman2011HR8799,  Madhusudhan2011HR8799, Skemer20112M1207, Marley2012HR8799} although their cloud properties are under discussion \citep{Barman2011HR8799, Barman20112M1207}. By contrast, the relatively blue $J-H$ (= --0.23) color of GJ 504 b is consistent with T-type brown dwarfs in the same temperature range \citep{Leggett2010LTdwarf}, which are representative of less cloudy atmospheres that could occur naturally in this temperature range \citep{Allard2001COND}. In addition, models examining cloud clearing as a function of temperature and surface gravity in \citet{Marley2012HR8799} do predict that a temperature of $\sim$500 K and $\log{g}$ of $\sim$4, as expected for GJ 504 b (see Table \ref{table: system_property}), should lead to a clear atmosphere, while the properties of previously imaged planets place them in a cloudy regime. The cold and perhaps less cloudy atmosphere of GJ 504 b thus places it in a physically distinct state from the hotter and cloudier atmospheres of previously imaged planets, and should be highly interesting for further atmospheric studies in the future. These properties imply that GJ 504 b will become a benchmark object for the study of exoplanet atmospheres. 

\subsection{Implications for the Origin of GJ 504 b}

GJ 504 b is a giant planet that lies at a projected separation of 43.5 AU. Two major scenarios could explain its formation: the core-accretion (CA) model \citep{Pollack1996CA} and the gravitational instability (GI) model \citep{Durisen2007GI}. Here, we discuss whether the CA or GI can account for the origin of GJ 504 b. \par

\subsubsection{Core Accretion}

In the CA model, the core of a giant planet accretes planetesimals and grows to a critical mass ($\sim$10 Earth masses), at which point the core begins to rapidly accrete gas directly from the protoplanetary disk. It is difficult for the CA model to explain the formation of giant planets in situ beyond 30 AU. A recent study \citep{Dodson-Robinson2009HR8799} suggested that even a locally stable (but globally unstable) disk with an order of magnitude higher mass than the minimum mass solar nebula \citep{Hayashi1981MMSN} around a G-type star like GJ 504 was unable, in simulations, to form giant planets beyond 30 AU within the lifetime of the protoplanetary disk \citep{Dodson-Robinson2009HR8799}. Therefore, the presence of GJ 504 b, as well as other directly-imaged planets such as HR 8799 bcde, remains a particular challenge for the conventional CA theory. In this implication, we should note such numerical investigations as conduced by \citet{Dodson-Robinson2009HR8799} have typically neglected some theoretical processes that may enable the cores of giant planets to grow in the shorter timescale: effective damping of small planetesimal's eccentricity due to gas drag \citep{Rafikov2004} or efficient capture of planetesimals due to atmospheric drag \citep{Inaba2003} may accelerate the accumulations of planetesimals in such the outer region, so that the in situ formation of GJ 504 b may become possible \citep{Rafikov2011}. However, even the models incorporating these processes may not necessarily be promising for producing a planet more massive than the critical mass beyond 30 AU at least for disks that are not particularly massive, as demonstrated by the simulations of \cite{Kobayashi2011}.\par

Even if the in situ formation of a giant planet is difficult in such outer regions, models incorporating planet scattering can account for the origin of GJ 504 b. If multiple planets are formed by CA in the inner regions, one or more of them could be scattered outward due to dynamical interactions among the planets \citep{Nagasawa2008scattering} or between the planets and the protoplanetary disk \citep{Paardekooper2010TypeI,Crida2009_common_gap}. GJ 504 b may have been pushed outward by as-yet undiscovered companions. Multiple giant planets may be common in metal-rich protoplanetary disks \citep{Fischer2005PlanetMetallicity}, because the high metallicity may enhance solid materials and planetesimals in their protoplanetary disks, resulting in rapid growth of planetary cores. Eventually, outward migration may frequently occur in the metal-rich systems. Hence, the relatively high metallicity of GJ 504 system (see Table. \ref{table: system_property}) may be a preferential property to explain the origin of GJ 504 b under the outward migration hypothesis. \par

\subsubsection{Gravitational Instability}

In contrast to CA, the GI model can form massive planets in situ at large radii. In this model, a massive protoplanetary disk becomes gravitationally unstable in its outer regions ($>$ $\sim$50 AU) as the cooling time becomes short relative to the local dynamical timescale \citep{Durisen2007GI}. The outer disk then fragments, collapsing directly into one or more giant planets. The instability can also occur between 20 and 50 AU, with continuous inflows of gas into the magnetically inactive region inducing fragmentation on a timescale of $\sim$10$^3$ yr \citep {Machida2011GI}. However, the planet would have to escape the very rapid inward migration expected in a turbulent disk \citep{Baruteau2011GImigration}. In addition, if multiple massive planets are formed in the GI disk, they would still be scattered and eventually ejected from the systems, as in the case of CA. Therefore, GJ 504 b would be a planet surviving against falling onto the central star or ejections form the system, if it was formed through the GI process. We should note that fragmentation could also be suppressed in a metal-rich disk like that expected for GJ 504; such a disk would cool less efficiently due to the higher opacities \citep{Gai2006GI}. Also, the significant gas accretion onto a clump formed according to the GI process may prevent the planet from remaining less massive \citep{Zhu2012, 2010Icar..207..509B}, disallowing the occurrence of a low-mass planet like GJ 504 b. 

\subsubsection{Summary for Discussions of the Origin of GJ 504 b and Future Prospect}

Based on the available data, we cannot conclusively determine which process led to the formation of GJ 504 b. However, it is interesting to note that GJ 504 b is the first wide-orbit giant planet detected around a demonstrably metal-rich star. As discussed above, the high metallicity may enhance massive planetary cores, while preventing the effective cooling of a protoplanetary disk, and thus be more hospitable for CA formation than GI formation. Thus, given these measurable properties, planets such as GJ 504 b may provide important clues for understanding the formation and evolution of giant planets. \par

For clarifying the origin of GJ 504 b, the further observations may be crucial. Intriguingly, the CA model suggests the presence of unseen massive planets at smaller angular separations \citep{Nagasawa2008scattering, Chatterjee2008Scattering}, making GJ 504 an outstanding candidate for follow-up observations by current and future instruments. $N$-body simulations for the dynamically unstable gas-free planetary system predict that such inner companions should orbit at a few AU from the host\footnote[8]{The current detection limit cannot explore giant planets in such the inner regions (see Figure \ref{fig: f3_contrast.eps}).} and have masses equal or greater than the corresponding outer planets \citep{Nagasawa2008scattering, Chatterjee2008Scattering}. Therefore, the inner counterpart for GJ 504 b would have to be more massive than $\sim$4 $M_{\rm{Jup}}$ if it exists, although we note that radial velocity observations have found few planets in this mass range at semimajor axes of a few AU \citep{Mayor2011HARPS} and the remaining gasses in the disk may affect the scattering of planets, changing the configurations of planets predicted by the simulations for the gas-free cases \citep{Moeckel2012Scattering}. \par

Also, further monitoring of the orbit may be important, since scattered planets preferentially have high eccentricities \citep{Chatterjee2008Scattering}. Furthermore, modeling the atmosphere of GJ 504 b may provide another piece of evidence to test whether it was produced by the CA process. The enhancement of heavy elements in the planetary atmosphere is a natural outcome of the CA process \citep{Chabrier2007}. It is possible to examine the abundances of such elements in the spectral energy distribution of GJ 504 b through the use of atmospheric models. For this purpose, multiple wavelength photometric observations as well as the spectroscopy can play an important role. \par

Further observations of the GJ 504 system will constrain the origin of GJ 504 b and may lead to a better understanding of the origin of planetary systems, including our solar System. Future instruments with direct (e.g., GPI, SPHERE, SCExAO) or indirect \citep[e.g., IRD;][]{Tamura2012IRD} techniques that can explore more inner region may be able to find a promising evidence to clarify the origin of GJ 504 b.

\section{Conclusion}

As part of the SEEDS direct-imaging survey, we have detected a giant planet around the Sun-like star GJ 504. The star has a spectral type of G0 and an approximate mass of 1.2 $M_{\rm{\odot}}$. Its age has been conservatively estimated as 160$^{+350}_{-60}$ Myr based on a combination of gyrochronology and chromospheric activity. Gyrochronology alone, which is the most direct and likely the most reliable age estimator, places the age at the lower end of this range, at 160$^{+70}_{-60}$ Myr. Observations over one year baseline provided confirmation that the detected planet is orbiting the star. Compared with previously imaged exoplanets, GJ 504 b has a number of interesting features.

\begin{enumerate}
\item The age of this planet is the oldest among all directly imaged planets. \vspace{-1.5mm}
\item Its mass is estimated to be 4$^{+4.5}_{-1.0}$ $M_{\rm{Jup}}$. This is among the least massive of imaged exoplanets. Because of its old age, the mass estimate is less dependent, compared to other imaged exoplanets, on the uncertainties of initial conditions in models used to convert its infrared luminosity into a mass.  \vspace{-1.5mm}
\item GJ 504 b is the first imaged giant planet on a wide orbit ($>$ 5 AU) around a G-type main-sequence star with a mass near the solar mass. \vspace{-1.5mm}
\item The planet's projected separation from the primary is about 43.5 AU, which is comparable to the widest-orbit planets around massive stars such as HR 8799. \vspace{-1.5mm}
\item The effective temperature of the planet (510$^{+30}_{-20}$ K) is the coldest among all previously detected giant planets.  \vspace{-1.5mm}
\item The blue $J-H$ color of --0.23 suggests that it has a less cloudy atmosphere than other imaged exoplanets. The cold temperature and color place the planet in a novel parameter space of atmospheres for exoplanets. \vspace{-1.5mm}
\item GJ 504 b is the first giant planet on a wide orbit discovered around a demonstrably metal-rich star.  \vspace{-1.5mm}

\end{enumerate} 

\vspace{-2mm}
The performed observations alone cannot conclusively uncover the origin of GJ 504 b. Further observations of this system will enable more direct comparisons with our own Solar System, and will help unveil the formation history of giant planets in the outer disk.
\vspace{-1.5mm}
\acknowledgments

The authors thank Dr. I. N. Reid for discussions about age estimations of main sequence stars. In addition, the authors have benefited from useful discussions with Drs. M. Ikoma, H. Nagahara, Y. Abe, S. Sugita, and B. Sato. Furthermore, the authors thank David Lafreni{\`e}re for generously providing the source code for the LOCI algorithm. Our referee's careful reading of the manuscript and excellent suggestions have improved the quality of this paper. The help from Subaru Telescope staff is greatly appreciated. This work is partly supported by KAKENHI 22000005 (M.T.), KAKENHI 23103002 (M.H.), WPI Initiative, MEXT, Japan (E.L.T), NSF AST 1008440 (C.A.G), NSF AST 1009203 (J.C.), and NSF AST 1009314 (J.P.W). Also, M.K. is partly supported by the Global COE program ``From the Earth to Earths". M.K., Y.H., Y.T., and J.K. are financially supported by the JSPS through the JSPS Research Fellowship for Young Scientists. Finally, the authors recognize and acknowledge the very significant cultural role and reverence that the summit of Mauna Kea has always had within the indigenous Hawaiian community. We are most fortunate to have the opportunity to conduct observations from this mountain.

{\it Facility:} \facility{Subaru (HiCIAO, IRCS)}

\appendix 


\section{Monte Carlo Orbital Analysis} 

\begin{figure}
\plotone{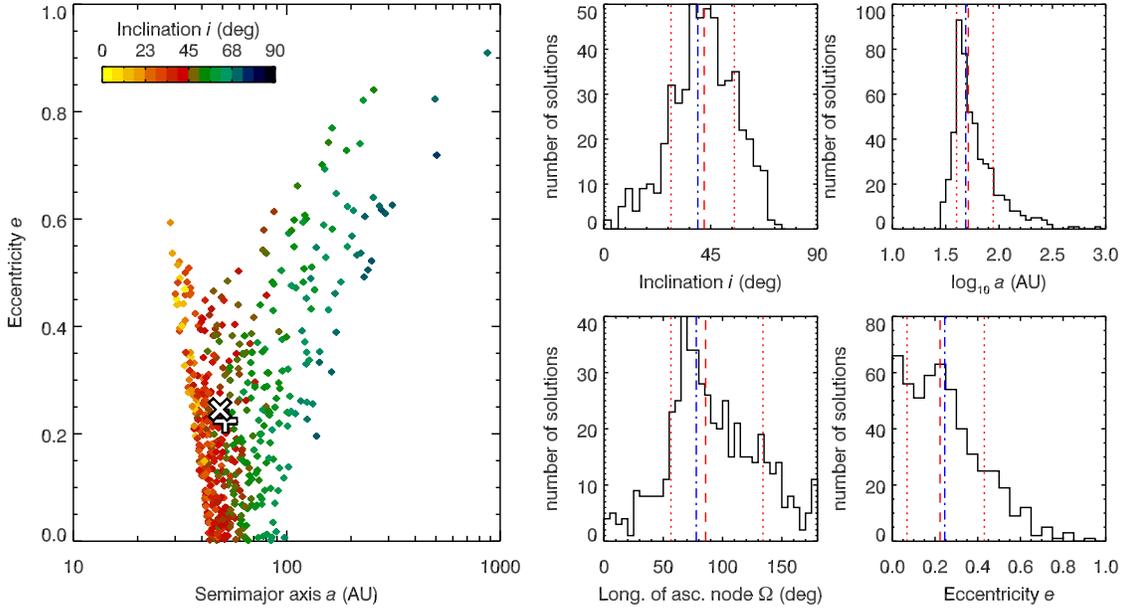}
\caption{Well-fitting family ($\chi^2 < \chi^2_{\rm{min}}+1$) of simulated orbits for GJ 504 b. In the $a$/$e$ scatter plot, a plus-sign marks the location of the weighted median, whereas a cross-sign marks the most likely orbit. In the histograms, red lines mark the weighted median (dashed) and the weighted 68\% interval (dotted), whereas a blue dash-dotted line marks the most likely orbit.}   
\label{fig: fig_appendix.eps}
\end{figure}

To demonstrate that the measured motion of GJ 504 b relative to its parent star GJ 504 is indeed consistent with the expected orbital motion of a gravitationally bound planet, we have run a Monte Carlo simulation following \citet{Janson2011GJ758B}. A large number of physically plausible orbits is randomly generated and compared to the astrometric data points. The maximum of the $\rm{\chi^2}$ function is typically ill-defined for an astrometric dataset covering only a small fraction of an orbit: an entire family of orbits spanning a large range of semimajor axes match the data equally well. Thus, rather than ascribing an undue amount of significance to the simulated orbit with the smallest $\rm{\chi^2}$ value, we consider the entire well-fitting family of orbits with $\rm{\chi^2} < \rm{\chi^2_{\rm{min}}}$ + 1. The plots in Figure \ref{fig: fig_appendix.eps} summarize the properties of this well-fitting orbit family.  \par

To characterize this distribution numerically, we determine the weighted median and the weighted 68\% range for each parameter of the orbital simulation (semimajor axis $a$, eccentricity $e$, inclination $i$, longitude of ascending node $\Omega$, and argument of periastron $\omega$). We calculate the statistical weight $W$ of each orbit as the mean orbital velocity divided by the local orbital velocity at the epoch of observation. This represents the fact that an eccentric planet orbits slower, spends more time, and is therefore more likely to be observed at apastron than at periastron. \par
	
However, some of the histograms are heavily skewed (e.g., $\log{a}$) or even periodic (e.g., $\Omega$); as a result, the median badly represents the true behavior of the entire well-fitting orbit family. For this reason, we further define the most likely orbit (MLO). \par

For each simulated orbit {\it n}, we calculate as a measure of likelihood $L_{\rm{n}}$ the product of the histogram values of the $\log{a}$, $e$, $i$, $\Omega$, $\omega$ histograms in the bin into which the orbit falls, as well as the statistical weight $W$: 

\begin{equation}
L_n = \rm{hist}_{\log{a}}(\log{a_n}) \cdot \rm{hist}_{e}(e_{n}) \cdot \rm{hist}_{i}(i_{n}) \cdot \rm{hist}_{\Omega}(\Omega_{n}) \cdot \rm{hist}_{\omega}(\omega_{n}) \cdot W.
\end{equation}

The MLO can then be defined as the orbit with the highest measure of likelihood, $L_{\rm{MLO}}$ = $\rm{max}_{n}$ $L_{n}$. 
  
\clearpage
\bibliographystyle{apj}

\clearpage

\end{document}